\documentclass[onecolumn]{article}

\usepackage[utf8]{inputenc}
\usepackage[big]{dgruyter}
\usepackage{graphicx}
\usepackage{amsmath, amsthm}

\usepackage{pdfpages}
\usepackage{booktabs}
\usepackage{graphicx}
\usepackage{comment}
\usepackage{multicol}
\usepackage{multirow}
\usepackage{color, colortbl}
\usepackage[title]{appendix}
\usepackage{float}
\usepackage{natbib}

\renewcommand{\thesection}{\arabic{section}}

\newtheorem{theorem}{Proposition}
\newtheorem{assumption}{Assumption}

\newcommand{\N}{\mathcal{N}}
  
\begin{document}

\articletype{Research Article{\hfill}Open Access}

\author*[1]{TingFang Lee}
\author[2]{Ashley L. Buchanan}
\author[3]{Natallia Katenka}
\author[4]{Laura Forastiere}
\author[5]{M. Elizabeth Halloran}
\author[7]{Georgios Nikolopoulos}

  \affil[1]{Department of Population Health, NYU Grossman School of Medicine; E-mail: Ting-Fang.Lee@nyulangone.org}

  \affil[2]{Department of Pharmacy Practice, University of Rhode Island; E-mail: buchanan@uri.edu}
  
  \affil[3]{Department of Computer Science and Statistics, University of Rhode Island; E-mail: nkatenka@uri.edu}
  
  \affil[4]{School of Public Health, Yale University; E-mail: laura.forastiere@yale.edu}
  
  \affil[5]{Biostatistics, Bioinformatics, and Epidemiology Program, Vaccine and Infectious Disease Division, Fred Hutchinson Cancer Center, and Department of Biostatistics, University of Washington; E-mail: betz@fredhutch.org}
  
  \affil[7]{Medical School, University of Cyprus; E-mail: nikolopoulos.georgios@ucy.ac.cy}

  \title{\huge Evaluating Spillover Effects in
Network-Based Studies In the Presence of
Missing Outcomes}

  \runningtitle{Spillover Effects in Network with Missing outcomes}

  \runningauthor{Lee et al.}
  

  \begin{abstract}
{Estimating causal effects in the presence of spillover among individuals embedded within a social network is often challenging with missing information. The spillover effect is the effect of an intervention if a participant is not exposed to the intervention themselves but is connected to intervention recipients in the network. In network-based studies, outcomes may be missing due to the administrative end of a study or participants being lost to follow-up due to study dropout, also known as censoring. We propose an inverse probability censoring weighted (IPCW) estimator, which is an extension of an IPW estimator for network-based observational studies to settings where the outcome is subject to possible censoring. We demonstrated that the proposed estimator was consistent and asymptotically normal. We also derived a closed-form estimator of the asymptotic variance estimator. We used the IPCW estimator to quantify the spillover effects in a network-based study of a nonrandomized intervention with possible censoring of the outcome. A simulation study was conducted to evaluate the finite-sample performance of the IPCW estimators. The simulation study demonstrated that the estimator performed well in finite samples when the sample size and number of connected subnetworks (components) were fairly large. We then employed the method to evaluate the spillover effects of community alerts on self-reported HIV risk behavior among people who inject drugs and their contacts in the Transmission Reduction Intervention Project (TRIP), 2013 to 2015, Athens, Greece. Community alerts were protective not only for the person who received the alert from the study but also among others in the network likely through information shared between participants. In this study, we found that the risk of HIV behavior was reduced by increasing the proportion of a participant's immediate contacts exposed to community alerts.}

\end{abstract}
  \keywords{Causal Inference, Interference/dissemination/spillover, Network studies, people who inject drugs, HIV/AIDS, Inverse probability weights}

  \journalname{Submitted to Journal of Causal Inference}
\DOI{DOI}
  \startpage{1}
  \received{..}
  \revised{..}
  \accepted{..}

  \journalyear{2021}
  \journalvolume{1}

\maketitle

\section{Introduction}

Estimating causal effects in the presence of spillover (dissemination or interference) among individuals in a network is often complicated by missing information. Consideration of missing outcome data is important when evaluating spillover within networks due to complex dependencies between individuals and ignoring missing data could lead not only to selection bias but also a distortion of the network structure. While there are many sources of missingness in the data (e.g., missing outcomes, baseline covariates, exposure status, network connections), we consider missing outcomes due to possible censoring, such as differential loss to follow-up due to study dropout, which can occur when participants who drop out differ from those who do not with respect to their exposure and outcome. One common approach for estimating of causal effects is to exclude all observations with missing outcomes, known as a complete case analysis. The complete case estimator is consistent if the observations are  missing completely at random (MCAR), meaning that missingness is unrelated to any measured variables. When the MCAR assumption does not hold and the missing at random assumption (MAR) is appropriate for the observed data, there are several approaches such as maximum likelihood, multiple imputation, fully Bayesian, and inverse probability weighting \cite{missing2005, seaman2011} that can be applied in this case.

Social networks can be represented as graphs displaying the connections among individuals, such as those defined by engagement in HIV risk behaviors (e.g., sexual/injection behaviors) among people who inject drugs (PWID). The exposure of one individual could influence the outcome of another individual, and this is known as interference \cite{tchetgen2012causal}. An effect of interest in this setting is the {\it spillover} (or indirect) effect, often defined as a contrast in average potential outcomes if an individual is unexposed, comparing different vectors of exposure for their spillover set (e.g., neighbors). In a network-based study, the spillover (or interference) set can be defined as the set of an individual’s neighbors (i.e., those who share a connection (edge or link) with that individual); therefore, the assumption of partial interference may not hold, which assumes \cite{hudgens2008toward, sobel2006randomized} that spillover is possible within a set or group of individuals, but not between groups. That is, no interference occurs between individuals in different groups but interference is possible between individuals within the same group defined by the interference set. In addition, the collection of interference sets is a partition of the study population. 

Inverse probability weighted (IPW) estimators for causal effects in the presence of interference in observational studies are often defined under the assumption of partial interference \cite{tchetgen2012causal}. Another approach defines interference by spatial proximity or network ties \cite{liu2016inverse, forastiere2016identification}, while allowing for overlapping interference sets. Liu et al. \cite{liu2016inverse} proposed an IPW estimator for a generalized interference set that allowed for overlapping between interference sets; however, the asymptotic variance was estimated under the assumption of partial interference defined by larger groupings or clusters. Forastiere et al. \cite{forastiere2016identification} quantified the effects through a subclassification estimator with a generalized propensity score, and a bootstrapping procedure with resampling at the individual-level or the cluster-level was used to quantify the variance. However, these approaches either rely on partial interference defined by larger clusters or resort to bootstrapping to derive estimators of the variance. In practice, ignoring the overlapping interference sets while estimating the variance can lead to inaccurate inference, and the resampling approaches can also be computationally intensive. To overcome these issues, Lee et al. \cite{nn_method} employed the IPW estimators from Liu et al. \cite{liu2016inverse} and Forastiere et al. \cite{forastiere2016identification} and proposed closed-form variance estimators while allowing for overlapping interference sets. Using statistical simulations, their proposed variance estimators were shown to be more efficient in network-based studies due to the use of additional information on connections between individuals. In addition to IPW approach, van der Laan \cite{vanderlaan2014} proposed a targeted maximum likelihood estimator to estimate the causal effects on longitudinal network studies. In their work, they aimed attention to particular types of causal quantities, namely the counterfactual mean under a stochastic intervention on the unit-specific treatment nodes. When the outcome of interest is subject to possible right censoring (i.e., events that occurred after a certain period time of follow-up are missing), Chakladar et al. \cite{halloran2019censoring}, and Loh et al. \cite{loh2018} extended IPW estimators under partial interference in an observational study to a setting with possible censoring using inverse probability of censoring weights (IPCW). In these approaches, the censoring weights are estimated using survival models, such as proportional hazards frailty models and accelerated failure time models. In addition, a clustering of observations is used to define the interference set (e.g., study clusters, geographic location) that allows for spillover within but not between clusters; however, the spillover between individuals with connections (i.e., edges) within a cluster or component, such as those observed due to HIV risk connections in a network-based study, were not considered \cite{aronow2017samii}. Liu et al. \cite{liu2016inverse} proposed generalized weighted type estimators with neighbor-level exposure weights that relaxed the partial interference assumption; however, this approach did not include censoring weights and the variance was estimated assuming partial interference. 

In this work, we develop methods to estimate causal effects in network-based observational studies in the presence of missing binary or continuous outcomes due to possible censoring. In this study, we employed a IPW estimator in Lee et al. \cite{nn_method}, which allows for spillover to each participant from their immediate connections in the network (i.e., neighbors) to quantify causal effects in network-based observational studies with a nonrandomized intervention. More precisely, each participant has a unique interference set determined by the observed network and interference sets can now overlap and no longer partition the network. We extend their IPW estimator to a setting with missing outcomes due to censoring using inverse probability censoring weights (IPCW), where we consider two different censoring mechanism: (1) censoring indicators are independent across participants conditional on baseline covariates and exposures or (2) censoring indicators are correlated between participants within a connected subnetwork or component in the network. Components are defined as subsets of connected participants, but not connected with participants in other components. We use the network structures including network connections between individuals and the independence of components to calculate a novel closed-form variance estimator by applying M-estimation. We evaluated finite-sample performance and the impact of the dependency assumption between censoring indicators of the proposed estimators in a simulation study.

We employed the proposed IPCW estimator to evaluate the direct and spillover effects of community alerts on self-reported HIV risk behavior at 6-month among PWID and their contacts in the Transmission Reduction Intervention Project (TRIP) from 2013 to 2015 in Athens, Greece \cite{nikolopoulos2016network, williams2018}. Previous work found evidence of a possibly meaningful spillover effect of community alerts on HIV risk behavior in TRIP; however, the prior study used a complete case analysis and did not consider the impact of missing outcomes on the validity of the analysis \cite{nn_method}.  PWID often participate in HIV risk behaviors that comprise HIV risk networks. In HIV risk networks, nodes represent individuals and edges between nodes represent two people who are engaging in HIV risk behavior, including sharing equipment for injection drug use or engaging in risky sexual behavior. In networks of PWIDs, interventions, including both educational and treatment-based, often have spillover causal effects, and overall intervention effects frequently depend on the network structure\cite{buchanan2018assessing, ghosh2017social}. 

The proposed IPCW is developed in Section 2. In Section 3, the results of a simulation study are reported that evaluates the performance of IPCW in finite-sample settings. In Section 4, we report the results that used the IPCW estimator to analyze the spillover effects of community alerts on self-reported HIV risk behavior among PWID in TRIP. We then discuss the performance of the IPCW estimators in the simulation study, limitation of the method, and future directions in Section 5.

\section{Methods}
In this section, we propose an IPCW estimator for network-based studies with binary or continuous outcomes subject to possible censoring due to differential loss to follow-up due to study drop out. If participants who drop out of the study differ from those who did not with respect to their outcome and exposure, then this may result in selection bias, even under the null \cite{robins2000selectionbias}. There are several types of missing mechanisms including missing completely at random (MCAR), missing at random (MAR), and missing not at random (MNAR). In the case of MCAR, the missing data mechanism is unrelated to any study variable. That is, the participants with completely observed data are assumed to be a random sample of all the participants assigned a particular intervention, which is often an unrealistic assumption in practice. MAR occurs when the missingness can be accounted for by fully observed study variables \cite{missingbook2014}. In other words, censoring indicators are independent conditional on baseline variables (e.g., covariates, exposures). MNAR assumes the missingness depends on underlying variables that we do not observe, such as participants with more severe substance use disorder are more likely to drop out of the study. In this paper, we focus on the MAR assumption, but the censoring models could be extended to some cases of MNAR \cite{mnar2013, mnar2018}.

The rest of this section is organized as follows: We first state the notation in Section 2.1 and the assumptions in Section 2.2. We then define the potential outcome framework and the population average causal effects of interest in Section 2.3. In Section 2.4, we define the inverse probability censoring weight estimator with censoring weights and neighbor-level exposure weights. The large sample variance estimators are derived using M-estimation theory \cite{stefanski2002boos} in Section 2.5.

\subsection{Notation}
\begin{figure}
\centering
    \includegraphics[scale=0.45]{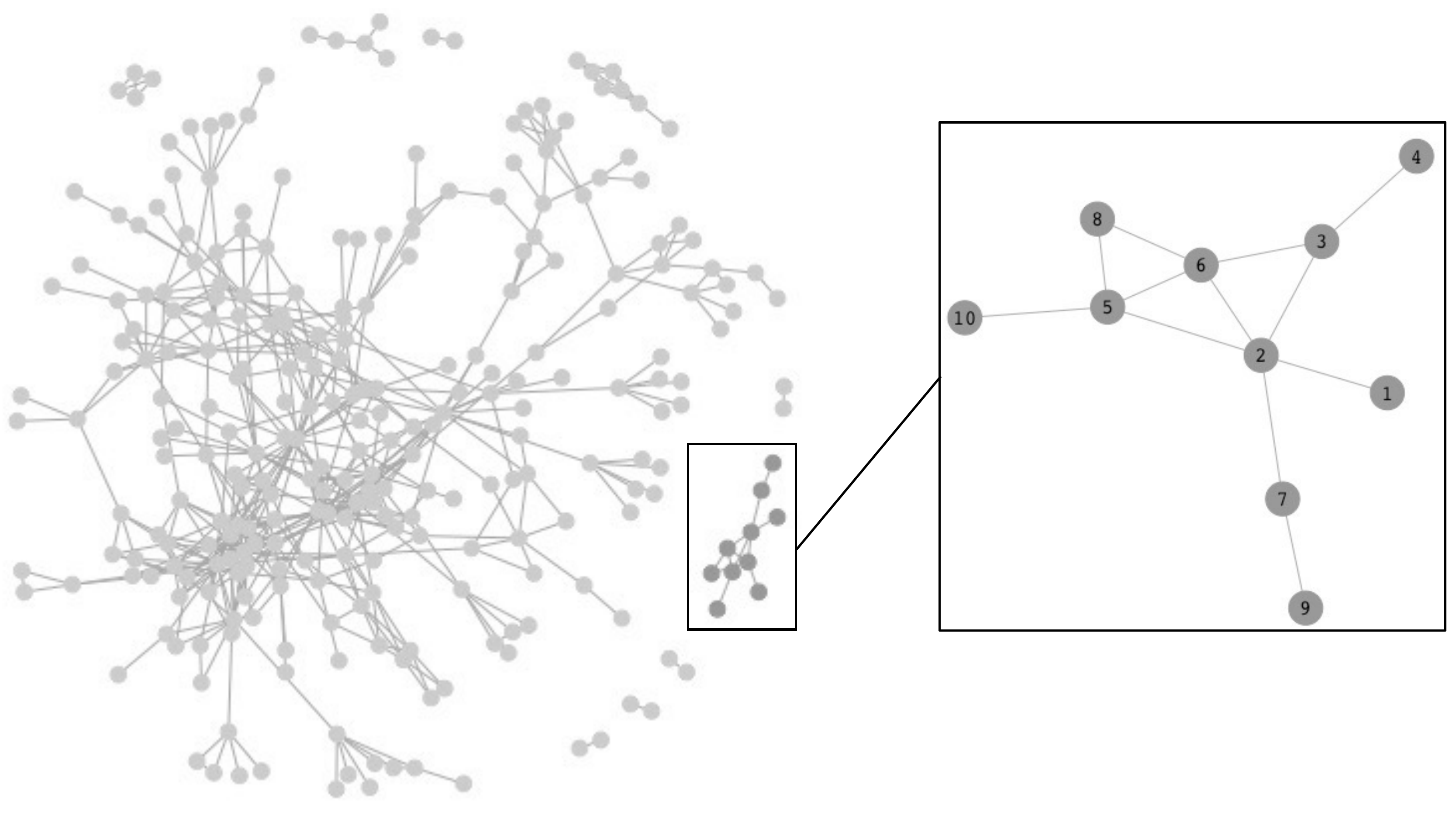}
    \caption{Left panel is the Athens PWID network. Right panel is enlarged one of the network components. }
    \label{fig:tripcomponents}
\end{figure}

A network is a structure consisting of a set of nodes, $V$, and a set of edges between nodes, $E$. Consider an HIV risk network, let $V=\{1, \hdots, n\}$ denote a set of participants in the study and define $E$ a set of the edges, where $e_{ij}=1$ if participants $i$ and $j$ engage in sexual/injection/drug use behavior together and $e_{ij}=0$ otherwise. Define the neighbors of participant $i$ by $\N_i=\lbrace j; e_{ij}=1 \rbrace$; in words, for each participant $i$, the neighbors are the immediate connections in the observed network excluding participant $i$. We denote $\N_i^*=\N_i\cup \{i\}$. Oftentimes, a network can be partitioned into connected subnetworks, $G_1, G_2, \cdots, G_m$ with $\nu= 1, \hdots, m$, each called components. There are no edges or connections between components, but possibly there are connections between participants within a component. Figure \ref{fig:tripcomponents} is the TRIP network with one of the network components enlarged. This component includes individuals labeled with $\{1, 2, 3, 4, 5, 6, 7, 8, 9, 10\}$, where, for example, the neighbors of node $6$ are $\{2, 3, 5, 8\}$ and the neighbors of node $2$ are $\{1, 3, 5, 6, 7\}$.

The degree of unit $i$ is denoted as $d_i = \sum_{j=1}^n e_{ij} =|\mathcal{N}_i|$. Let $A_i$ be the binary baseline exposure of participant $i$ with $A_i=1$ if exposed and 0 otherwise. Let $Z_i$ denote the vector of baseline pre-exposure covariates for participant $i$. We denote the vector of intervention exposures for the neighbors for participant $i$ as  ${A}_{\mathcal{N}_i}= \lbrace A_{ ij}; e_{ij}=1 \rbrace$.  We also denote the vector of baseline pre-exposure covariates for the neighbors for participant $i$ as  ${Z}_{\mathcal{N}_i}= \lbrace Z_{ ij}; e_{ij}=1 \rbrace$. Denote realizations of $A_i$ by $a_i$ and $Z_i$ by $z_i$. 

\subsection{Assumptions}
The proposed approach requires identification assumptions in this observational network design.  

\begin{assumption}
\label{ass:nint}
The neighbor interference assumption \citep{forastiere2016identification} assumes that the outcome of participant $i$ only depend on their exposure and the exposures of their neighbors $\N_i$ and not on the exposures of others besides their neighbors in the network. By consistency, $$Y_i=y_i(A_i, A_{\N_i}).$$
\end{assumption}
Let $y_i(a_i, {a}_{\mathcal{N}_i})$ denote the potential outcome of individual $i$ if they received exposure $a_i$ and their neighbors received ${a}_{\mathcal{N}_i}$. Therefore, the potential outcome of participant $i$ depends not only on their own exposure, but also on the exposures of their neighbors. Let $Y_i = y_i(A_i, {A}_{\mathcal{N}_i})$ denote the observed outcome, which holds by (causal) consistency. For example, in Figure \ref{fig:tripcomponents}, $Y_7=Y_7(a_7, (a_2, a_9))$, which is that the outcome for individual 7 is affected by their own exposure and the exposure of individual 2 and 9 only and no other individuals' exposures in either the component or network.

\begin{assumption}\label{assump:irrelavance}
We assume that the treatment or intervention assignment mechanism does not affect the outcome. More precisely, if there are different versions of the intervention, we assume that those are irrelevant for the causal contrasts of interest and that we have one version of intervention and one version of no intervention \citep{forastiere2016identification}. 
\end{assumption}

There are no multiple versions of this exposure, and if there were multiple versions of exposure, those are assumed to be irrelevant for the causal effect of interest \cite{vanderweele2009irrelavance}.

\begin{assumption} The conditional exchangeability assumption \cite{nn_method},
$$y(a_i, a_{\N_i}) \perp A_i, A_{\N_i} |Z_i, Z_{\N_i}.$$
\end{assumption}

Assumption 3 means that exposure to the intervention is independent of potential outcomes conditional on baseline pre-exposure covariates of participant $i$ and their neighbors.

\begin{assumption}\label{assump:positive}
Positivity assumption for the intervention exposure \cite{liu2016inverse},
$\Pr(A_i=a_i|Z_i=z_i)>0$ and $\Pr(A_i=a_i, A_{\N_i}=a_{\N_i}|Z_i=z_i, Z_{\N_i}=z_{\N_i})>0$ for all $a_i$, $a_{\N_i}$, $z_i$, and $z_{\N_i}$.
\end{assumption}

Positivity assumes that each participant has a non-zero probability of being assigned every possible intervention exposure status given every combination for the levels of the observed covariates.

\begin{assumption}
 Conditional independent censoring $C_i  \perp y_i(a_i, a_{\N_i})|Z_i, Z_{\N_i}, A_i, A_{\N_i}$.
\end{assumption}

Assumption 4 states that the censoring of participant $i$ is independent of their potential outcomes conditional on their own exposure and pre-exposure covariates and their neighbors' exposures and pre-exposure covariates.

\begin{assumption} (Conditional exposure independence) Conditional on the exposures and covariates for an individual's neighbors and the neighborhood-level random effect, the exposure $A_i$ for individual $i$ and the exposure $A_j$ for individual $j$ are independent. That is, given the neighbors' exposures $A_{\N_i}$ and $A_{\N_j}$, neighbors' covariates $Z_{\N_i}$ and $Z_{\N_j}$, and neighborhood-level random effect $b_{\N_i^*}$ and $b_{\N_j^*}$,  
$$A_i \perp A_j| A_{\N_i}, Z_{\N_i}, b_{\N_i^*}, A_{\N_j}, Z_{\N_j}, b_{\N_j^*}.$$
\end{assumption}
The nearest neighbor-level random effect $b_{\N_i^*}$ accounts for possible correlation of exposures among individual $i$ and their neighbors $\N_i$. 

\subsection{Spillover Parameters}


In this study, we use the network neighbors $\N_i$ of an individual as the spillover set for that individual. Let $\alpha$ represents the counterfactual scenario in which participants in $a_{\N_i}$ are exposed with probability $\alpha$ and we refer to this parameter as the \textit{intervention} coverage among the neighbors. We do not assume that $A_1, \hdots, A_n$ are independent Bernoulli random variables; however, this distribution of exposure assignment is used to define the counterfactuals. Let $\pi(a_{\N_i};\alpha)=\alpha^{\sum a_{\N_i}}(1-\alpha)^{d_i-\sum a_{\N_i}}$ denote the probability of the neighbors of individual $i$ receiving exposure $a_{\N_i}$ under allocation strategy $\alpha$ where $d_i=|\N_i|$. Let $\pi(a_i;\alpha)=\alpha^{a_i}(1-\alpha)^{1-a_i}$ denote the probability of individual $i$ receiving exposure $a_i$ and $\pi(a_i, a_{\N_i};\alpha)=\pi(a_{\N_i};\alpha)\pi(a_i;\alpha)$ denote the probability of individual $i$ together with their neighbors receiving the set of exposures $(a_i, a_{\N_i} )$.

Recall that $y_i(a_i, a_{\N_i})$ denote the potential outcomes for participant $i$ under individual exposure $a_i$ and network neighbors' exposure $a_{\N_i}$. The average potential outcome under allocation strategy $\alpha$ is defined by 
\begin{equation}
    \bar{y}(a, \alpha)=\frac{1}{n}\sum_{i=1}^n \sum_{a_{\N_i}}y_i(a_i=a, a_{\N_i})\pi(a_{\N_i};\alpha). \label{eqn:po1}
\end{equation} The marginal average potential outcome is
\begin{equation}
    \bar{y}(\alpha)=\frac{1}{n}\sum_{i=1}^n \sum_{a_i, a_{\N_i}}y_i(a_i, a_{\N_i})\pi(a_i, a_{\N_i};\alpha).\label{eqn:po2}
\end{equation}

Different contrasts of these population average causal effects are often of interest in public health research. In particular, spillover effects can assess the public health impact of an intervention in the network among individuals who were not exposed to intervention but who are connected to recipients. We defined these effects under allocation strategies $\alpha, \alpha_0, \alpha_1$ as follows. The direct effect is $\overline{DE}(\alpha)=\bar{y}(1, \alpha)-\bar{y}(0, \alpha)$, which is a difference in the risk of the outcome when a participant is exposed to the intervention compared to when a participant is not exposed  with exposure coverage level $\alpha$ for the neighbors. The spillover or indirect effect is $\overline{IE}(\alpha_1, \alpha_0)=\bar{y}(0, \alpha_1)-\bar{y}(0, \alpha_0)$, which is a difference in the risk of the outcome if a participant is not exposed to the intervention under two different intervention coverage levels for the neighbors. The total effect is defined as $\overline{TE}(\alpha_1, \alpha_0)=\bar{y}(1, \alpha_1)-\bar{y}(0, \alpha_0)$, which is a measure of the maximal intervention effect, and is a difference in the risk if a participant is exposed under one intervention coverage level for the neighbors $\alpha_1$ compared if a participant is unexposed under intervention coverage level for the neighbors $\alpha_0$. The overall effect is $\overline{OE}(\alpha_1, \alpha_0)=\bar{y}(\alpha_1)-\bar{y}(\alpha_0)$, which is the difference in average potential outcomes under one coverage for participant $i$ (index) and their neighbors compared to another coverage for the participant $i$ and the neighbors. The effects can also be considered on a ratio scale, such as a risk ratio $\overline{DE}(\alpha)=\bar{y}(1, \alpha)/\bar{y}(0, \alpha).$ In this study, we will focus on the estimation of risk differences. 

According to Assumption \ref{ass:nint}, spillover is assumed to be possible to a participant from their neighbors in the network (i.e., a participant's immediate or first-degree contacts). The spillover set or interference set of participant $i$ is defined as the set of their neighbors. Assumption 1 relaxes the partial interference assumption which implies no interference between individuals in different groups but allow for possible interference between individuals within the same group, requiring that individuals only belong to one interference set \cite{tchetgen2012causal, hudgens2008toward}. Assumption 1 allows for participants to be part of more than one interference set using information on connections in the network. For example, in Figure \ref{fig:tripcomponents}, the interference set of participant 3 is $\{2, 4, 6\}$ and the interference set of participant 5 is $\{2, 6, 8, 10\}$.  Participant 3 and 5 are not neighbors but the interference sets overlap and both include participants $\{2, 6\}$.

\subsection{Inverse Probability of Censoring Weighted (IPCW) Estimator}
In a network-based study, data are ascertained from a network with $m$ distinct components, where components are connected subnetworks, $G_1, G_2, \hdots, G_m$, in the full network with $\nu=1,\cdots,m$. Suppose the outcomes are subject to possible censoring, e.g., due to loss to follow-up. Let $C_i$ be the binary censoring indicator for participant $i$ where $C_i=1$ if an individual is censored (i.e., missing outcome) and 0 otherwise (i.e., outcome observed). We consider two alternative censoring mechanisms:
\begin{itemize}
        \item[i.] Given the baseline pre-exposure covariates and exposure, the censoring indicators $C_i$ are independent across participants; that is, for participants $i$ and $j$, $C_i \perp C_j |Z_i, Z_{\N_i}, A_i, A_{\N_i}, Z_j, Z_{\N_j}, A_j, A_{\N_j}$.
        \item[ii.] There is a correlation between the censoring indicators of participants within a component. That is, if $i, j \in G_\nu$, then $C_i \perp C_j |Z_i, Z_{\N_i}, A_i, A_{\N_i}, Z_j, Z_{\N_j}, A_j, A_{\N_j}, \rho_\nu$ where $\rho_\nu\sim N(0, \gamma^2)$, for some $\gamma$, is the component-level random effect. Otherwise if $i, j \not\in G_\nu$, $C_i\perp C_j |Z_i, Z_{\N_i}, A_i, A_{\N_i}, Z_j, Z_{\N_j}, A_j, A_{\N_j}$.  
\end{itemize}
 Censoring mechanism (i) states the censoring of person $i$ conditional on their own and their neighbors' covariates and exposures is independent of the censoring of person $j$ conditional on their own and their neighbors' covariates and exposure. Censoring mechanism (ii) states the censoring of person $i$ conditional on their own and their neighbors' covariates and exposures as well as the component-level random effect is independent of the censoring of person $j$ conditional on their own and their neighbors' covariates, exposure, and the component-level random effect.\footnote{More precisely, we consider that $C_i\perp C_j$ given their own and their neighbors' exposure status and covariates. That is $Pr(C_i, C_j|Z_i, Z_{\N_i}, A_i, A_{\N_i}, Z_j, Z_{\N_j}, A_j, A_{\N_j})=Pr(C_i|Z_i, Z_{\N_i}, A_i, A_{\N_i}, Z_j, Z_{\N_j}, A_j, A_{\N_j})\cdot Pr(C_j|Z_i, Z_{\N_i}, A_i, A_{\N_i}, Z_j, Z_{\N_j}, A_j, A_{\N_j})=Pr(C_i|Z_i, Z_{\N_i}, A_i, A_{\N_i})\cdot P(C_j|Z_j, Z_{\N_j}, A_j, A_{\N_j})$.} These considerations may be plausible in network intervention studies in the scenario that participants may choose to return for study visits or not, regardless of their neighbors' visit attendance. Under different assumptions of the censoring mechanisms, we will use different methods to model the conditional probability of censoring. In our approach, regardless of the model for the censoring mechanism, we assume that the outcomes are missing at random; that is, adjustment for missingness can be accounted for using measured baseline covariates and the exposure status to estimate a censoring weight. 

The conditional probability of censoring can be estimated using a logistic model when considering censoring mechanism(i); 
\begin{align}
    S_C(C_i|Z_i, Z_{\N_i}, A_i, A_{\N_i})&=Pr(C_i=0|Z_i, Z_{\N_i}, A_i, A_{\N_i})=1-Pr(C_i=1|Z_i, Z_{\N_i}, A_i, A_{\N_i})\notag\\
    &=1-{\rm logit}^{-1}( (Z_i, Z_{\N_i})\cdot \xi_z+(A_i, A_{\N_i})\cdot\xi_a).\notag
\end{align}

Alternatively, censoring mechanism (ii) considers that there is correlation of the censoring mechanisms between participants within a component. In this case, a mixed-effects model can be used to model the censoring mechanism: 
$$S_C(C_i|Z_i, Z_{\N_i}, A_i, A_{\N_i}, \rho_\nu)=1-Pr(C_i=1|Z_i, Z_{\N_i}, A_i, A_{\N_i}, \rho_\nu)=1-{\rm logit}^{-1}((Z_i, Z_{\N_i})\cdot \xi_z+(A_i, A_{\N_i})\cdot\xi_a+\rho_\nu),$$ where individual $i$ belongs to component $G_\nu$ and $\rho_\nu\sim N(0, \gamma^2)$ for some $\gamma$. In other words, the conditional probability of censoring is the probability that the outcome of individual $i$ will not be observed at the follow-up visit.

We used a mixed-effects model to estimate the exposure propensity score for each participant and their interference set $\N_i$: 
$$f(A_i, A_{\N_i}|Z_i, Z_{\N_i})=\int \prod_{j\in\N_i^*} p_j^{A_j}(1-p_j)^{1-A_j}f(b_{\N_i^*}; 0, \psi)db_{\N_i^*},$$ where $\N_i^*=\N_i\cup \{i\}$ and $i \in G_{\nu}$,  $p_j=\Pr(A_j=1|Z_j, b_{\N_i^*})=\mbox{logit}^{-1}(Z_j\cdot \theta_z+b_{\N_i^*})$, and $f(b_{\N_i^*}; 0, \psi)\sim N(0, \psi)$. Here, $b_{\N_i^*}$ is the nearest neighbors-level random effect accounting for possible correlation of exposures among individual $i$ and their neighbors $\N_i$.

Under Assumption 1-6, an extension of the nearest neighbor IPW estimator \citep{nn_method} of the study population average outcome for exposure $a$ and intervention allocation strategy $\alpha$ that now includes an inverse probability of censoring weight is defined as
$$\widehat{Y}^{IPCW}(a, \alpha)=\frac{1}{n}\sum_{i=1}^n \frac{y_i(A_i, A_{\N_i})I(C_i=0)I(A_i=a)\pi(A_{\N_i};\alpha)}{f(A_i, A_{\N_i}|Z_i, Z_{\N_i})S_C(C_i|Z_i, A_i)}.$$ An estimator of the population average marginal outcome for intervention allocation strategy $\alpha$ is defined as 
$$\widehat{Y}^{IPCW}(\alpha)=\frac{1}{n}\sum_{i=1}^n \frac{y_i(A_i, A_{\N_i})I(C_i=0)\pi(A_i, A_{\N_i};\alpha)}{f(A_i, A_{\N_i}|Z_i, Z_{\N_i})S_C(C_i|Z_i, A_i)}.$$

\noindent Under allocation strategies $\alpha$, $\alpha_0$, and $\alpha_1$, we consider the following risk difference estimators of the direct, spillover (indirect), composite (total), and overall effects:
\begin{align}
    &\widehat{DE}(\alpha)=\widehat{Y}^{IPCW}(1, \alpha)-\widehat{Y}^{IPCW}(0, \alpha)\notag\\
    &\widehat{IE}(\alpha_1, \alpha_0)=\widehat{Y}^{IPCW}(0, \alpha_1)-\widehat{Y}^{IPCW}(0, \alpha_0)\notag\\
    &\widehat{TE}(\alpha_1, \alpha_0)=\widehat{Y}^{IPCW}(1, \alpha_1)-\widehat{Y}^{IPCW}(0, \alpha_0)\notag\\
    &\widehat{OE}(\alpha_1, \alpha_0)=\widehat{Y}^{IPCW}(\alpha_1)-\widehat{Y}^{IPCW}( \alpha_0).\notag
\end{align}

\begin{theorem}
If the exposure propensity scores and censoring weights are known, then the IPCW estimator is unbiased: $E[\widehat{Y}^{IPCW}(a, \alpha)]=\bar{y}(a, \alpha)$ and $E[\widehat{Y}^{IPCW}( \alpha)]=\bar{y}(\alpha)$.
\end{theorem}
The proof of this proposition is in Appendix A. We can easily get unbiased estimators for the causal effects because the causal effects are different contrasts of the average potential outcomes.

\subsection{Large Sample Properties of the IPCW Estimator}
The large sample variance estimators can be derived using M-estimation theory \cite{stefanski2002boos}. We assume that an observed network can be expressed as the union of connected subnetworks, referred to as components. Given a network with $n$ participants and $m$ components $\{G_1, G_2, \cdots, G_m\}$, let $Y_\nu=\{Y_i|i \in V(G_\nu)\}$, $A_\nu=\{A_i|i \in V(G_\nu)\}$, $Z_\nu=\{Z_i|i \in V(G_\nu)\}$, and $C_\nu=\{C_i|i \in V(G_\nu)\}$ where $V(G_\nu)$ is the set of nodes in $G_\nu$ with $\nu=1, \hdots, m$. The observable random variables $(Y_\nu, A_\nu, Z_\nu, C_\nu)$ for $\nu=1, \hdots, m$ are assumed to be independent but not necessarily identically distributed. We assume that the $m$ components are a
random sample from the infinite super-population of groups and the size of each component is bounded \cite{stefanski2002boos}.

We denote $\Theta=\{\theta_z, b_{\N_i^*}\}$ the set of coefficients in the exposure propensity score, including the fixed effects and the random effect, and by $\Phi$ the set of coefficients in the censoring weight model including fixed effect coefficients and random effect standard deviation, respectively. 
We derive the results in this section under the assumption of correlated censoring indicators. In this case, the censoring model is a mixed effects model, and let $\Phi=\{\xi_z, \xi_a, \rho_\nu\}$ represent the parameters, including fixed effects coefficients and random effect standard deviations. If the censoring model is a logistic model assuming independent censoring indicators, $\Phi=\{\xi_z, \xi_a\}$ represents the parameters including fixed effect coefficients in the censoring model and the procedure would otherwise be the same as below. 

Let $Y_{G_\nu}=(Y_{0, G_\nu}, Y_{1, G_\nu}, Y_{2, G_\nu})$ represent the sum of the potential outcomes in component $G_v$ under allocation strategy $\alpha$, 
\begin{align}
    Y_{0, G_\nu}&=\sum_{j\in V(G_\nu)}\sum_{a_{\N_j}} y_j(a_j=0, a_{\N_j})\pi(a_{\N_j};\alpha)\notag\\
     Y_{1, G_\nu}&=\sum_{j\in V(G_\nu)}\sum_{a_{\N_j}} y_j(a_j=1, a_{\N_j})\pi(a_{\N_j};\alpha)\notag\\
      Y_{2, G_\nu}&=\sum_{j\in V(G_\nu)}\sum_{a_j, a_{\N_j}} y_j(a_j, a_{\N_j})\pi(a_j, a_{\N_j};\alpha).\notag
\end{align}

We use $m$ independent components, while preserving the underlying connections of an individual’s neighbors comprising the network structure of each component to conduct inference. That is, by extending Liu, Hudgens and Becker-Dreps \cite{liu2016inverse}, every individual is now assigned their own propensity score based on the observed network structure defined by their neighbors. Whereas in Liu, Hudgens and Becker-Dreps \cite{liu2016inverse}, statistical inference was conducted by assuming partial interference in which the study population was partitioned into non-overlapping groups and all individuals in a group were assigned one group-level propensity score. Let $\theta_{0, \alpha}=\bar{y}(0, \alpha)=\sum_{\nu=1}^m Y_{0, G_\nu}/n$, $\theta_{1, \alpha}=\bar{y}(1, \alpha)=\sum_{\nu=1}^m Y_{1, G_\nu}/n$, and $\theta_\alpha=\bar{y}(\alpha)=\sum_{\nu=1}^m Y_{2, G_\nu}/n$. Let $k=E[|V(G_\nu)|]$ be the average component size in the population.

Given the set of observed variables $(Y_\nu, A_\nu, Z_\nu, C_\nu)$,
the maximum likelihood estimators of the coefficients $\Theta$ and $\Phi$ and can be written as as system of estimating equations $\displaystyle\sum_{\nu=1}^m \psi(Y_\nu, A_\nu, Z_\nu, C_\nu; \Theta, \Phi)=0$, where $\psi(Y_\nu, A_\nu, Z_\nu, C_\nu; \Theta, \Phi)$ is a vector of 
estimating equations $\psi=(\psi_{\gamma}, \psi_{\eta}, \psi_1, \psi_2, \psi_3)^T$ as defined below:
\begin{align}
    &\psi(Y_\nu, A_\nu, Z_\nu, C_\nu; \gamma)=\frac{1}{k}\sum_{j\in V(G_\nu)}\frac{\partial \log f(A_j, A_{\N_j}|Z_j, Z_{\N_j})}{\partial \gamma}, \gamma\in \Theta\notag\\
    &\psi(Y_\nu, A_\nu, Z_\nu, C_\nu; \eta)=\frac{1}{k}\sum_{j\in V(G_\nu)}\frac{\partial \log S_C(C_j|Z_j, A_j)}{\partial \eta}, \eta\in\Phi\notag\\
    &\psi_0(Y_\nu, A_\nu, Z_\nu, C_\nu;\alpha, \theta_{0, \alpha})=\frac{1}{k}\sum_{j\in V(G_\nu)}\frac{y_j(A_j, A_{\N_j})I(C_j=0)I(A_j=0)\pi(A_{\N_j};\alpha)}{f(A_j, A_{\N_j}|Z_j, Z_{\N_j})S_C(C_j|Z_j, A_j)}-\theta_{1, \alpha}\notag\\
    &\psi_1(Y_\nu, A_\nu, Z_\nu, C_\nu;\alpha, \theta_{1, \alpha})=\frac{1}{k}\sum_{j\in V(G_\nu)}\frac{y_j(A_j, A_{\N_j})I(C_j=0)I(A_j=1)\pi(A_{\N_j};\alpha)}{f(A_j, A_{\N_j}|Z_j, Z_{\N_j})S_C(C_j|Z_j, A_j)}-\theta_{0, \alpha}\notag\\
    &\psi_2(Y_\nu, A_\nu, Z_\nu, C_\nu;\alpha, \theta_\alpha)=\frac{1}{k}\sum_{j\in V(G_\nu)}\frac{y_j(A_j, A_{\N_j})I(C_j=0)\pi(A_j, A_{\N_j};\alpha)}{f(A_j, A_{\N_j}|Z_j, Z_{\N_j})S_C(C_j|Z_j, A_j)}-\theta_\alpha\notag.
\end{align}

Let $\theta=(\Theta, \Phi, \theta_{0, \alpha}, \theta_{1, \alpha}, \theta_{\alpha})$, and $$\psi_\nu(Y_\nu, A_\nu, Z_\nu, C_\nu; \alpha, \theta)=\displaystyle\begin{pmatrix}
    \psi(Y_\nu, A_\nu, Z_\nu, C_\nu; \gamma)\\
    \psi(Y_\nu, A_\nu, Z_\nu, C_\nu; \eta)\\
    \psi_0(Y_\nu, A_\nu, Z_\nu, C_\nu;\alpha, \theta_{0, \alpha})\\
    \psi_1(Y_\nu, A_\nu, Z_\nu, C_\nu;\alpha, \theta_{1, \alpha})\\
    \psi_2(Y_\nu, A_\nu, Z_\nu, C_\nu;\alpha, \theta_\alpha)
\end{pmatrix}_{\gamma\in\Theta, \eta\in\Psi}$$ such that $\displaystyle\sum_{\nu=1}^m\psi_\nu(Y_\nu, A_\nu, Z_\nu, C_\nu; \alpha, \hat{\theta})=0$. Note that $\hat{\theta}$ is the solution for $\theta$ for this vector of estimating equations.

Under suitable regularity conditions that as $m\rightarrow\infty$, the closed form sandwich type estimator of the variance is a consistent and asymptotically normal by Proposition 2 below. Let $A(\theta)=E[-\dot{\psi}(Y_\nu, A_\nu, Z_\nu, C_\nu; \alpha, \hat{\theta})]$, and $B(\theta)=E[\psi(Y_\nu, A_\nu, Z_\nu, C_\nu; \alpha, \hat{\theta})\psi(Y_\nu, A_\nu, Z_\nu, C_\nu; \alpha, \hat{\theta})^T]$.
\begin{theorem}
Under suitable regularity conditions and due to the unbiased estimating equations,
  $\sqrt{m}(\hat{\theta}-\theta)$ converges in distribution to $N(0, \Sigma_m)$ as $m\rightarrow\infty$ where the covariance matrix is given by $\displaystyle \Sigma_m=\frac{1}{m}A^{-1}(\theta)B(\theta)(A^{-1}(\theta))^{\rm T}.$
\end{theorem}

More details of Proposition 2 can be found in Appendix A. A consistent estimator of the variance for the IPCW estimator is given in Appendix A. This variance estimator can be used to construct Wald-type confidence intervals (CIs) for the direct, spillover, total, and overall effects.

\section{Simulation}
A simulation study was conducted to evaluate the finite-sample performance of the IPCW estimators and their corresponding closed-form variance estimators. We focused on evaluating the finite sample bias and coverage of the corresponding 95\% Wald-type confidence intervals. The observed network was generated as following steps:
\begin{itemize}
    \item[Step 1: ] Generate $m$ regular network components of degree four, where the number of nodes is sampled from a Poisson distribution with an average number of nodes of 10.
    \item[Step 2: ] Define each of these $m$ networks to constitute a single component of the observed network.
\end{itemize}
We conducted the experiments varying the number of components, $m\in\{10, 50, 100, 200\}$, and the size of each component was 10 on average across all simulation runs. Given a generated network, a total of 1,000 data sets were simulated in the following these steps. The network characteristics (number of components, number of nodes in each component), parameters in the outcome model, and censoring models were informed by estimates in the TRIP data. Although we assumed that the censoring indicators are conditionally independent given their own and their neighbors' covariates and exposures, we did not find significant associations between the censoring indicators and neighbors' covariates, their own and their neighbors' exposures in TRIP analysis. Therefore, we do not include them in our simulation study. The data was generated in the following steps:
\begin{itemize}
    \item[1.] Generate a baseline covariate $Z_i\sim \mbox{Bern}(0.5)$.
    \item[2.] Assign the random effects for the censoring model, $\rho_\nu\sim N(0, 0.3^2)$, and random effects for the component-level exposure propensity score $b_\nu$ and $b_\nu\sim N(0, 0.5^2)$, $\nu=1, 2, ..., m$.
    \item[3.] Given the covariate $Z_i$ and assuming that the censoring mechanism are independent across participants, the censoring indicator is $$C_i\sim \mbox{Bern}(\mbox{logit}^{-1}(-3+2\cdot Z_i)).$$ Under the assumption that there is correlation between participants within a component, the censoring indicator is $$C_i\sim \mbox{Bern}(\mbox{logit}^{-1}(-3+2\cdot Z_i+\rho_\nu)).$$
    \item[4.] We then generate the potential outcomes 
    $$y_i(a_i, a_{\N_i})\sim \mbox{Bern}(\mbox{logit}^{-1}(-1.75+0.5\cdot a_i + \frac{\sum a_{\N_i}}{d_i}-1.5 a_i\cdot \frac{\sum a_{\N_i}}{d_i} +0.5 Z_i)).$$
    \item[5.] The observed exposures are generated as $$A_i\sim \mbox{Bern}(\mbox{logit}^{-1}(0.7-1.4\cdot Z_i+b_\nu))$$ where $i \in V(G_\nu)$.
    \item[6.] Based on the observed exposures generated in Step 5, the observed outcome for each individual $i$ is $Y_i=y_i(A_i, A_{\N_i})$. 
\end{itemize}

For each simulated data set, the $\widehat{Y}^{IPCW}(a, \alpha)$ and $\widehat{Y}^{IPCW}( \alpha)$ are evaluated for $a = 0, 1$ and $\alpha = \lbrace 0.25, 0.5, 0.75 \rbrace$. The true parameters were calculated by averaging the generated potential outcomes as in equations (\ref{eqn:po1}) and  (\ref{eqn:po2}). The estimated standard errors (ASE) were derived from the estimated variance matrix $\hat{\Sigma}_m$ in Appendix A and were averaged across all simulations. The ASE in each data set were used to obtain corresponding Wald 95\% confidence intervals for each simulation. Empirical standard errors (ESE) were the standard deviation of estimated averaged across all simulated data sets. Empirical coverage probability (ECP) was defined as the proportion of simulated data sets that the true parameter was within the 95\% confidence interval among the total 1000 simulations.

\begin{figure}[t]
\centering
\caption{The average absolute value of bias (left) and empirical coverage probability (ECP) (right) on networks with 10, 50, 100, and 200 components using logistic regression censoring model (top) and mixed-effects censoring model (bottom)}
\includegraphics[scale=0.55]{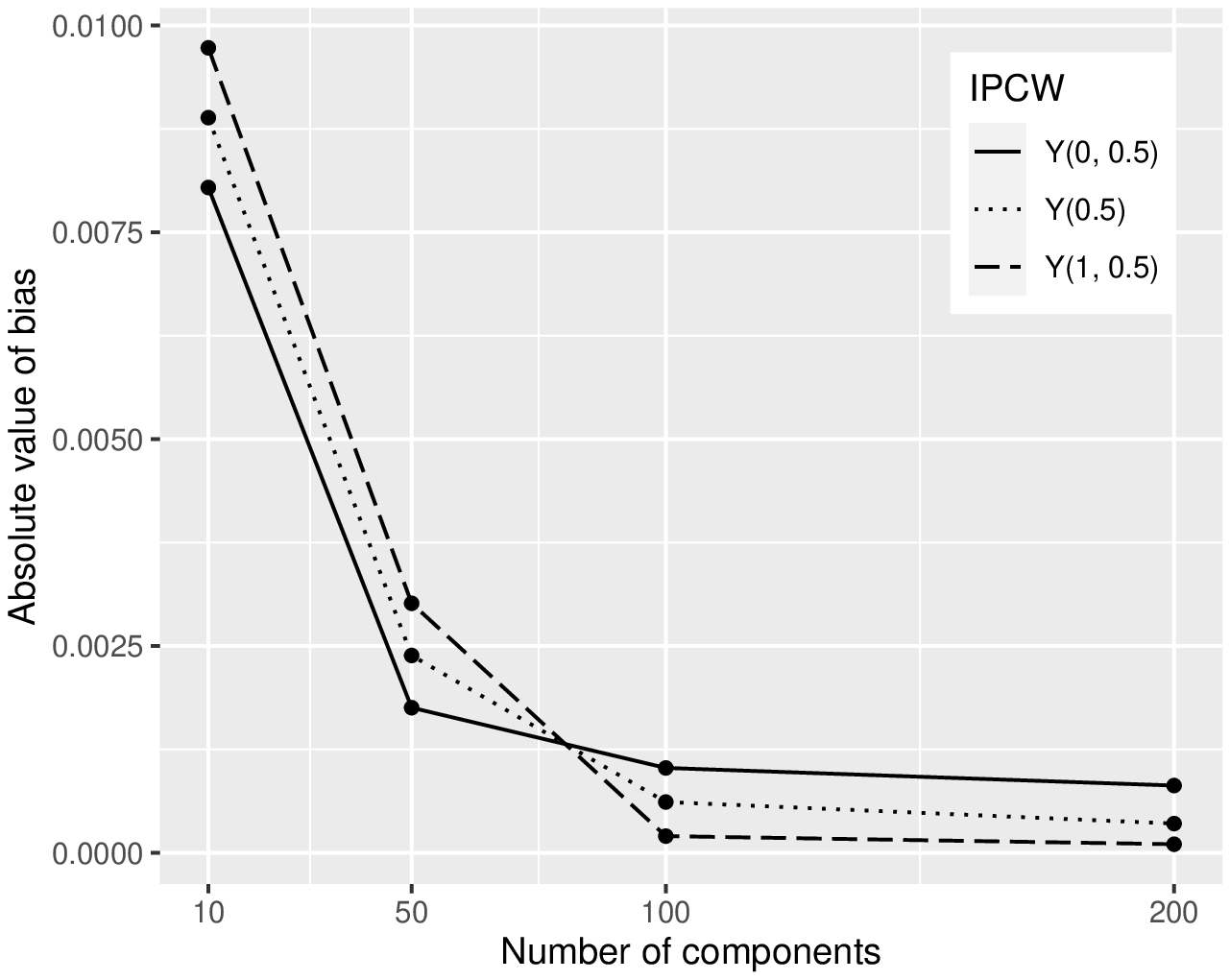}\includegraphics[scale=0.55]{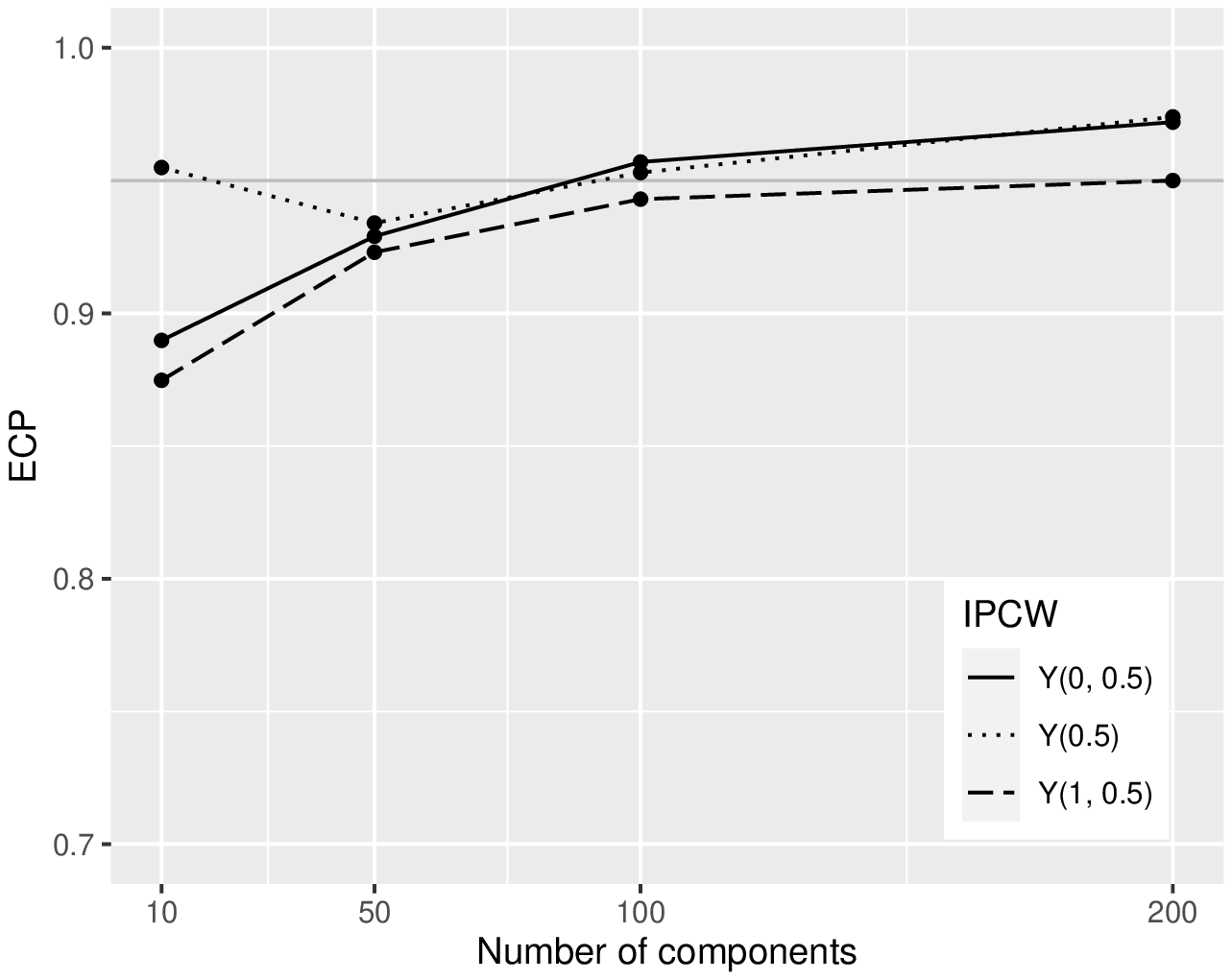}
\includegraphics[scale=0.55]{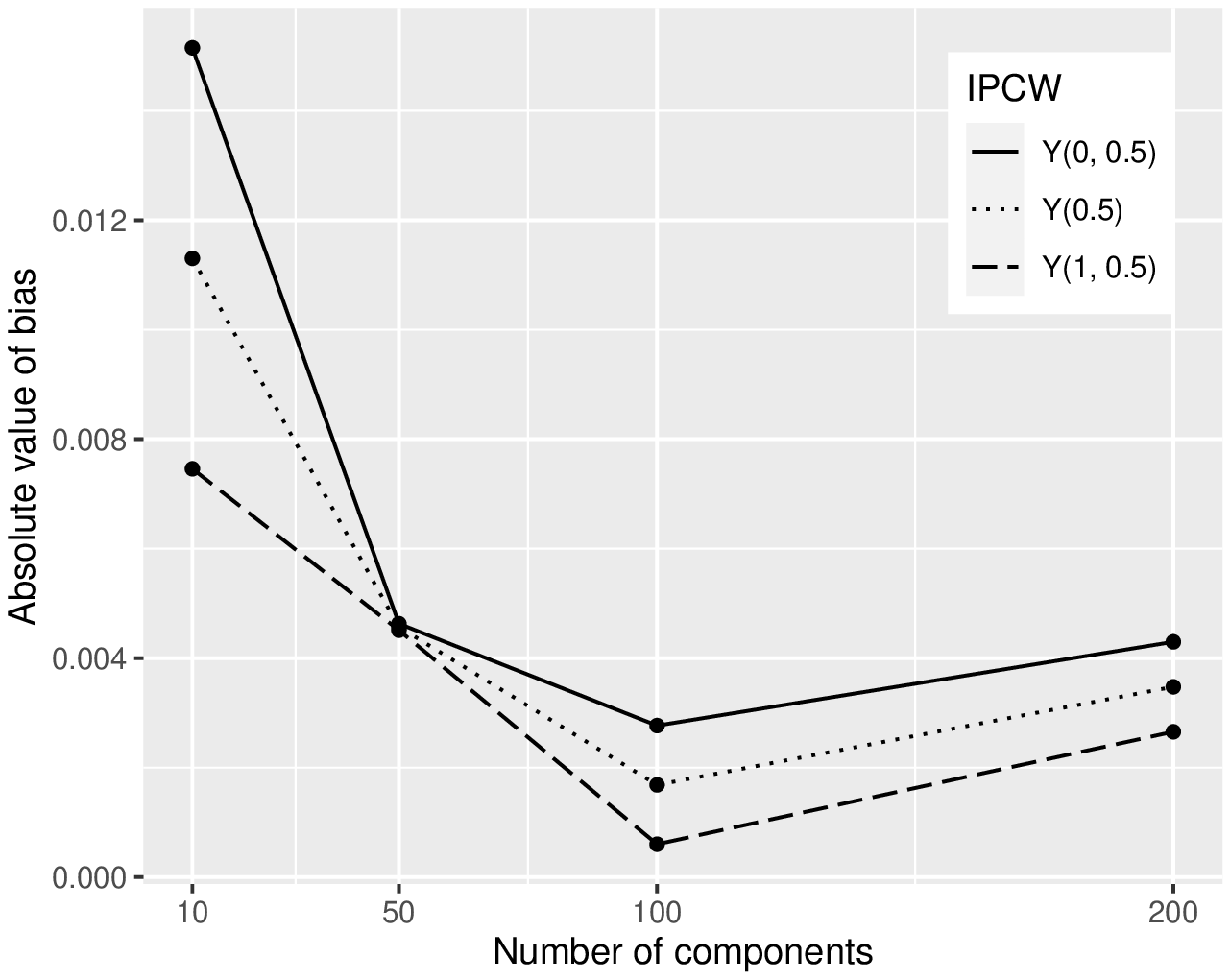}\includegraphics[scale=0.55]{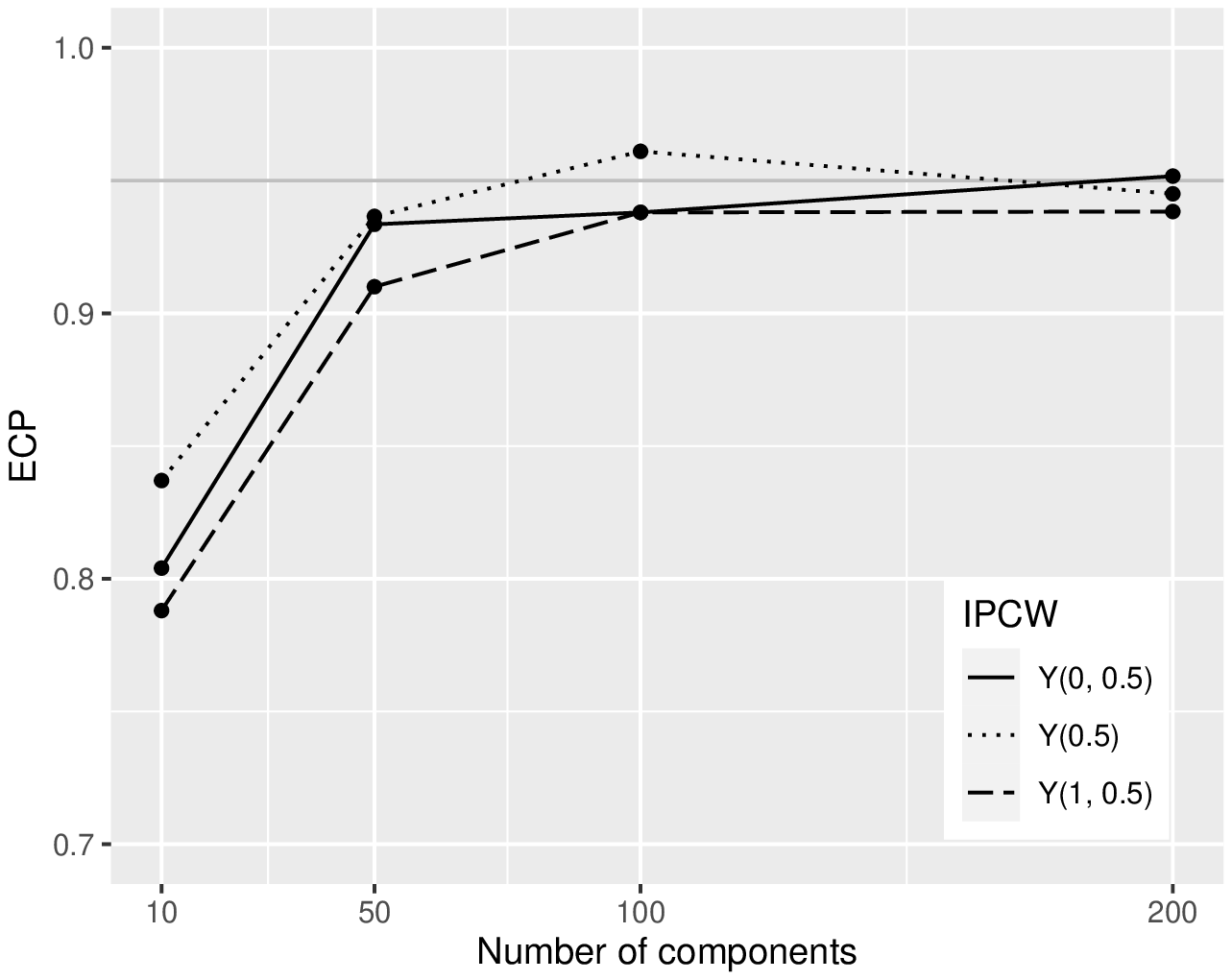}
\label{fig:sim_result}
\end{figure}

We compared the performance of the estimators for networks with 10, 50, 100, and 200 components. Figure \ref{fig:sim_result} illustrated the absolute value of average bias and ECPs of $\hat{Y}(1, 0.5)$, $\hat{Y}(0, 0.5)$, and $\hat{Y}(0.5)$. Although the absolute value of bias for networks with 100 components is slightly smaller ($<0.002$) than for networks with 200 components when considering the mixed effects censoring models with 50\% allocation strategy, the absolute value of bias for 200 components network under allocation strategies 25\% and 75\% were smaller than for 100 components network (Table \ref{tab:result_200} and \ref{tab:result_100}). We conclude that, in general, the absolute value of average bias decreases and the empirical coverage probability approaches 95\% as the number of components increases. In Table \ref{tab:result_200}, we report the simulation results of network with 200 components. The absolute values of average bias over different allocation strategies are less than $0.003$, and the empirical coverage probabilities are around the nominal value of 95\%.

\noindent In addition to the main simulation scenarios above, we also considered the following scenarios:
\begin{itemize}
    \item[1.] We evaluated the impact on the performance of the estimators when there was a higher level of dependency between censoring indicators for the mixed-effects model approach. We used the same simulated regular network with degree 4 and 100 components to compare the performance of the average outcome estimators. The censoring indicators were generated by $$C_i\sim \mbox{Bern}(\mbox{logit}^{-1}(-3+2\cdot Z_i +\rho_\nu)).$$ We increased the complexity of dependency by increasing the uncertainty of censoring random effects $\rho_\nu\sim N(0, r^2)$, where $r=0.1, 0.2, 0.3, 0.4, 0.5$. The results demonstrated that the performance of both IPCW estimators were comparable when the variance of censoring random effects increased. Summarized results can be found in Appendix C.
    \item[2.] We considered the network structure from our motivating study TRIP. The TRIP network consisted of 10 components with 277 nodes and 542 edges. Based on the main simulation scenarios results, a small number of components may result in poor finite-sample performance of variance estimators. In addition, if there is substantial variation in the observed component sizes, the average component size used to derive the estimator of the variance may not be appropriate unless the average potential outcomes in the component are independent of component size \cite{liu2016inverse}. To increase the number of components based on network structure and generate components that were more comparable in size for estimation of the variance of the estimated causal effects, we employed an efficient modularity-based, fast greedy, approach to detect communities to further divide some large connected components of the TRIP network into a total of 24 smaller and denser components. By ignoring sparser edges between components, we treated the obtained communities as independent units to possibly improve the estimation of the variance with more components similar in size. Importantly, we still defined the interference sets using the neighbors for point estimation of the causal effects. The results illustrate that the variance estimators were conservative on original TRIP network (10 components) and anti-conservative on the 24 components network under both censoring mechanisms (Table \ref{tab:trip_sim_logit} and \ref{tab:trip_sim_mix}). 
\end{itemize}

\begin{table}[t]
  \centering
  \caption{Simulation results from 1000 simulated datasets of IPCW estimators using logistic regression censoring model on original TRIP network (10 components) (left) and the network divided by community detection into 24 components (right).}
    \begin{tabular}{lrrr|cc|cc}
    \toprule
         &      &      &      & \multicolumn{2}{c|}{10 components} & \multicolumn{2}{c}{24 components} \\
         & \multicolumn{1}{c}{True} & \multicolumn{1}{c}{Bias} & \multicolumn{1}{c|}{ESE} & \multicolumn{1}{c}{ASE} & \multicolumn{1}{c|}{ECP} & \multicolumn{1}{c}{ASE} & \multicolumn{1}{c}{ECP} \\
    \midrule
    \midrule
    Y(1, 0.25) & 0.2488 & 0.0130 & 0.121 & 0.196 & 0.970  & 0.088 & 0.872 \\
    Y(1, 0.5) & 0.2269 & 0.0037 & 0.078 & 0.182 & 0.998 &  0.070 & 0.934 \\
    Y(1, 0.75) & 0.2051 & -0.0070 & 0.145 & 0.176 & 0.969 &  0.076 & 0.904 \\
    Y(0, 0.25) & 0.2288 & -0.0045 & 0.121 & 0.203 & 0.988 &  0.078 & 0.922 \\
    Y(0, 0.5) & 0.2761 & 0.0039 & 0.078 & 0.226 & 1.000 &  0.079 & 0.956 \\
    Y(0, 0.75) & 0.3262 & 0.0050 & 0.241 & 0.265 & 0.983 &  0.109 & 0.886 \\
    Y(0.25) & 0.2338 & -0.0002 & 0.096 & 0.199 & 0.998 &  0.070 & 0.937 \\
    Y(0.5) & 0.2515 & 0.0038 & 0.053 & 0.203 & 1.000 &  0.064 & 0.982 \\
    Y(0.75) & 0.2354 & -0.0040 & 0.121 & 0.196 & 0.996 &  0.073 & 0.945 \\
    \bottomrule
    \end{tabular}%
    
    {\footnotesize ESE: empirical standard error; ASE: asymptotic standard error; ECP: empirical coverage probability.}
  \label{tab:trip_sim_logit}%
\end{table}

\begin{table}[h]
  \centering
  \caption{Simulation results from 1000 simulated datasets of IPCW estimators using mixed effects censoring model on original TRIP network (10 components) (left) and the network divided by community detection into 24 components (right). }
    \begin{tabular}{lrrr|cc|cc}
    \toprule
         &      &      &      & \multicolumn{2}{c|}{10 components} & \multicolumn{2}{c}{24 components} \\
         & \multicolumn{1}{c}{True} & \multicolumn{1}{c}{Bias} & \multicolumn{1}{c|}{ESE} & \multicolumn{1}{c}{ASE} & \multicolumn{1}{c|}{ECP} & \multicolumn{1}{c}{ASE} & \multicolumn{1}{c}{ECP} \\
    \midrule
    \midrule
    Y(1, 0.25) & 0.2492 & 0.0139 & 0.125 & 0.229 & 0.960 & 0.090 & 0.859 \\
    Y(1, 0.5) & 0.2276 & 0.0056 & 0.074 & 0.203 & 0.996 & 0.069 & 0.928 \\
    Y(1, 0.75) & 0.2066 & -0.0032 & 0.128 & 0.188 & 0.966 & 0.073 & 0.883 \\
    Y(0, 0.25) & 0.2296 & -0.0022 & 0.113 & 0.327 & 0.973 & 0.076 & 0.891 \\
    Y(0, 0.5) & 0.2796 & 0.0090 & 0.073 & 0.314 & 0.996 & 0.080 & 0.941 \\
    Y(0, 0.75) & 0.3270 & 0.0224 & 0.116 & 0.328 & 0.968 & 0.109 & 0.870 \\
    Y(0.25) & 0.2345 & 0.0018 & 0.091 & 0.301 & 0.994 & 0.069 & 0.913 \\
    Y(0.5) & 0.2523 & 0.0073 & 0.050 & 0.257 & 0.999 & 0.064 & 0.975 \\
    Y(0.75) & 0.2367 & 0.0032 & 0.098 & 0.221 & 0.992 & 0.068 & 0.925 \\
    \bottomrule
    \end{tabular}%

     {\footnotesize ESE: empirical standard error; ASE: asymptotic standard error; ECP: empirical coverage probability.}
  \label{tab:trip_sim_mix}%
\end{table}%

\section{Motivating study: The Transmission Reduction Intervention Project}
Participants in the Transmission Reduction Intervention Project (TRIP) study were people who inject drugs (PWID) and their contacts who lived in Athens, Greece between 2013 and 2015. PWID who participated in the ARISTOTLE project at HIV testing centers in Athens were initially recruited into the TRIP study if they were found to be recently infected with HIV. ARISTOTLE was a community-based programme aiming to mitigate HIV transmission among PWID by implementing a care system that involved reaching out to high-risk PWID, engaging them in HIV testing, and initiating HIV care, opioid substitution treatment, and antiretroviral therapy \cite{sypsa2014, aristotle}. In TRIP, each newly diagnosed individual was asked to identify their recent sexual and drug use partners in the last six months. These partners were then recruited and asked to identify their sexual and drug use partners, who were also recruited and contacts to other individuals already recruited in the study were also ascertained. If any of these partners were determined to be recently infected with HIV, then their contacts and the contacts of their contacts (i.e., two waves of contact tracing) were recruited as well, including their possible connections to the other participants in the study. Complete details about the TRIP study design can be found in previously published papers \cite{nikolopoulos2016network, nikolopoulos2017evaluation}.

In addition to HIV testing, the study provided access to treatment as prevention (TasP), referrals for medical care, and distributed community alerts to inform community members about temporary increases in the risk for HIV acquisition. For example, a community alert would be distributed to those in close proximity in the observed network to a recently-infected participant. These alerts included paper flyers given to participants and posted in a location frequented by members of the local PWID community. We considered those who received the alerts from the study staff or flyers to be exposed to the community alert, while the remaining participants were not exposed but could have possibly received this information from their exposed neighbors. All participants completed computer-assisted interviews and also had their HIV status ascertained. They provided demographic information, answered questions about engagement in risk behaviors, HIV status, substance use, access to care, HIV knowledge, stigma, injection norms, and their opinions on the project. Follow-up interviews were conducted with participants about six months after they completed their baseline interview \cite{nikolopoulos2016network}.

\begin{table}[t]
  \centering
  \caption{Descriptive statistics of network characteristics and baseline variables for study participants after excluding isolates in the Transmission Reduction Intervention Project, Athens, Greece, 2013-2015.}
    \begin{tabular}{rrrr}
    \toprule
    \multicolumn{1}{l}{\multirow{4}[2]{*}{Network Characteristic}} & \multicolumn{2}{r}{Nodes} & 275 \\
         & \multicolumn{2}{r}{Edges} & 540 \\
         & \multicolumn{2}{r}{Average Degree (SD)} & 3.9 (3.5) \\
         & \multicolumn{2}{r}{Density} & 0.014 \\
    \midrule
    \multicolumn{1}{l}{\multirow{2}[2]{*}{Community alert}} & \multicolumn{2}{r}{Exposed} & 29 (11\%) \\
         & \multicolumn{2}{r}{Not Exposed} & 246 (89\%) \\
    \midrule
    \multicolumn{1}{l}{\multirow{2}[2]{*}{HIV status}} & \multicolumn{2}{r}{Positive } & 142 (52\%) \\
         & \multicolumn{2}{r}{Negative} & 133 (48\%) \\
    \midrule
    \multicolumn{1}{l}{\multirow{2}[2]{*}{Date of first interview}} & \multicolumn{2}{r}{Before ARISTOTLE ended} & 130 (47\%) \\
         & \multicolumn{2}{r}{After ARISTOTLE ended} & 145 (53\%) \\
    \midrule
    \multicolumn{1}{l}{\multirow{4}[2]{*}{Education}} & \multicolumn{2}{r}{Primary School or less} & 87 (32\%) \\
         & \multicolumn{2}{r}{High School (first 3 years)} & 82 (30\%) \\
         & \multicolumn{2}{r}{High School (last 3 years)} & 68 (25\%) \\
         & \multicolumn{2}{r}{Post High School} & 38 (13\%) \\
    \midrule
    \multicolumn{1}{l}{\multirow{4}[2]{*}{Employment}} & \multicolumn{2}{r}{Employed} & 44 (16\%) \\
         & \multicolumn{2}{r}{Unemployed; looking for work} & 64 (23\%) \\
         & \multicolumn{2}{r}{Can't work; health reason} & 128 (47\%) \\
         & \multicolumn{2}{r}{Other} & 39 (14\%) \\
    \midrule
    \multicolumn{1}{l}{Shared injection equipment } & \multicolumn{2}{r}{Yes} & 207 (75\%) \\
    \multicolumn{1}{l}{in last 6 months} & \multicolumn{2}{r}{No} &  68 (25\%)\\
    \midrule
    & \multirow{2}[1]{*}{\hspace{1cm}Yes} & Exposed & 11 (4\%) \\
         &      & Not Exposed &  83 (30\%)\\
    \multicolumn{1}{l}{Outcome: sharing injection} & \multirow{2}[1]{*}{\hspace{1cm}No} & Exposed & 14 (5\%) \\
    \multicolumn{1}{l}{equipment at the 6-month visit}   &      & Not Exposed &  111 (40\%)\\
     & \multirow{2}[1]{*}{\hspace{1cm}Missing} & Exposed & 4 (1\%) \\
         &      & Not Exposed &  52 (20\%)\\
    \bottomrule
    \end{tabular}%
  \label{tab:addlabel}%
\end{table}%

We applied the IPCW estimators to evaluate the causal effects of community alerts on HIV risk behavior ascertained by the report of risk behavior at the six-month follow-up visit. The exposure community alerts was defined as receiving a study flyer, either from study staff or viewing one posted in the community. We determined the outcome of HIV risk behavior ascertained at the six-month visit as a self-report of shared drug equipment (e.g., needles, syringes) in the last six months. We consider the report of any injection HIV risk behavior at a 6-month visit as a binary outcome. The following pre-exposure baseline covariates are included in the adjusted models for both the exposure and censoring based on expert knowledge and were known or suspected risk factors for the outcome: HIV status, shared drug equipment (i.e. self-report of sharing or being shared drug equipment (e.g., needles, syringes)) in the last six months prior to baseline, the calendar date of the first interview (binary: before or after ARISTOLE program ended), education (primary school, high school, and post-high school), and employment status (employed, unemployed/looking for a job, cannot work because of health reason, and others). The causal effects were estimated separately using the logistic regression censoring model under the assumption that censoring mechanisms are independent across participants and the mixed-effects censoring model under the assumption that there is a correlation between censoring of participants within a component. The results showed that the participant's, their neighbors' exposure to community alerts, and neighbors' baseline covariates were not statistically significant associated with censoring indicators.

The network structure in TRIP included 356 participants and 542 shared connections. One of the participant was recruited twice as a network member of a recent seed and as a network member of a control seeds with long-term HIV infection. In our analysis, we only used the information for this participant corresponding to their records as a network member of a recent seed. 79 participants were isolates (i.e. not sharing connection with other network members) and removed for our analysis as spillover is not possible for isolates. In addition, 2 participants were removed due to missing values on HIV risk behavior in the past 6 months reported at baseline. The final TRIP network had 10 unique components (component sizes were 2, 2, 2, 2, 2, 4, 5, 7, 10, 239) with 275 participants and 540 shared connections among those individuals after excluding isolates. There were 56 participants (21\%) who were lost to follow-up by six months. Among the 275 participants in TRIP, 29 participants (11\%) received a community alert about an increased risk for HIV infection. The point estimates and corresponding 95\% Wald-type confidence intervals use both censoring models under allocation strategies 25\%, 50\%, and 75\%, representing low, moderate and high coverage strategies. The mixed-effects censoring model did not detect random effects in each component on the original TRIP network. Therefore, the results using either censoring model are identical (Table \ref{tab:TRIP_10}).

\begin{table}[htbp]
  \centering
  \caption{The estimated risk difference (RD) and 95\% Wald-type confidence intervals (CI) of the effects of community alerts at baseline on HIV risk behavior at 6 months on the original TRIP network (10 components) (left) and estimated using completed cases ($n=216$) (right) under allocation strategies 25\%, 50\%, and 75\%.}
    \begin{tabular}{lccccc}
    \toprule
    \multirow{2}[2]{*}{Effects} & Coverage & \multicolumn{2}{c}{Censoring Model} & \multicolumn{2}{c}{Complete cases}\\
          & ($\alpha$, $\alpha'$) & RD  & 95\% CI & RD  & 95\% CI \\
         
    \midrule
    \midrule
     Direct & (25\%, 25\%) & -0.1228 & (-0.288, 0.042) & -0.0772 & (-0.181, 0.027)\\
     Direct & (50\%, 50\%) & -0.2299 & (-0.637, 0.177) & -0.1761 & (-0.489, 0.137)\\
     Direct & (75\%, 75\%) & -0.1741 & (-0.514, 0.166) &-0.2393	& (-0.700, 0.222) \\
    Indirect & (50\%, 25\%) & -0.0565 & (-0.117, 0.004) &-0.0404	 & (-0.081, 0.001) \\
    Indirect & (75\%, 50\%) & -0.1502 & (-0.382, 0.081) & -0.0818	 & (-0.186, 0.023) \\
    Indirect & (75\%, 25\%) & -0.2066 & (-0.494, 0.081)& -0.1222	& (-0.262, 0.018)\\
     Total & (50\%, 25\%) & -0.2864 & (-0.747, 0.174) & -0.2165 &	(-0.551, 0.118) \\
     Total & (75\%, 50\%) & -0.3243 & (-0.894, 0.246) &-0.3211 & (-0.873, 0.231) \\
     Total & (75\%, 25\%) & -0.3808 & (-1.004, 0.242) & -0.3615 & (-0.937, 0.214)\\
    Overall & (50\%, 25\%) & -0.1407 & (-0.358, 0.076) & -0.1091 & (-0.266, 0.048) \\
    Overall & (75\%, 50\%) & -0.1658 & (-0.447, 0.116) & -0.1732 & (-0.458, 0.111) \\
    Overall & (75\%, 25\%) & -0.3065 & (-0.805, 0.192) & -0.2824 & (-0.723, 0.158)\\
    \bottomrule
    \end{tabular}%
  \label{tab:TRIP_10}%
\end{table}%

These results indicate that the risk of HIV behavior was reduced by increasing the proportion of a participant's neighbors exposed to community alerts, in addition to a participant's exposure. The estimated direct effect under allocation strategy 75\% was $-0.17$ (CI: $-0.51, 0.17$); that is, we would expect 17 fewer reports of risk behavior per 100 participants if an individual receives alert compared to if an individual did not receive an alert with 75\% intervention coverage (i.e., 75\% of their neighbors receiving alerts). The spillover effect under allocation strategies 25\% versus 75\% was $-0.21$ (CI: $-0.50, 0.08$). We would expect 21 fewer reports of risk behavior per 100 participants if a participant does not receive an alert with 75\% intervention coverage compared to only 25\% intervention coverage.  The total effect under allocation strategies 75\% versus 25\% was $-0.38$ (CI: $-1, -0.24$); that is, we expect 38 fewer reports of risk behavior per 100 participants if an individual receives the alert and 75\% of their neighbors also receive an alert compared to if an individual does not receive an alert and only 25\% of their neighbors receive an alert. The overall effect was $-0.31$ (CI: $-0.81, 0.19$), indicating we expect 31 fewer reports of risk behavior per 100 participants if 75\% of the neighbors and the index participant receives alert compared to only 25\% of the neighbors and index participant receive alerts. Interestingly, the estimated spillover effect was larger in magnitude than the estimated direct effect, possibly due to the information about the community alert already being available in the spillover set due to the intervention coverage level of 75\%. The estimated spillover effect is a meaningful reduction in the report of risk behavior with a 21\% reduction from a baseline prevalence of 75\%.  

To improve the validity of the estimation of the asymptotic variance of the causal effect estimates, we use an efficient modularity-based (e.g., fast greedy) algorithm \cite{PhysRevE.70.066111} to detect communities to further divide the TRIP network into 24 components. Results are summarized in Appendix D (Table \ref{tab:trip}). The 95\% CI estimates using 24 components were narrower than the analysis using 10 components. These results are aligned with the additional simulation scenario 1 and 2 in the simulation study. 

\section{Conclusion}
In this paper, we extended the neighbor inverse probability weighted estimator \cite{nn_method} to allow for possible censoring of outcomes in network-based studies. The two IPCW estimators were obtained by including a censoring weight derived from two different censoring models, namely logistic regression and mixed-effects model, and a inverse probability weighted estimator that assumes the interference set comprises the first-degree connections for each participants. The additional simulation scenario of varying variances of random effects in the censoring weighted model suggested that the performance in terms of empirical coverage probabilities of both censoring models, with and without correlations of censoring indicators, was comparable. The main difference between the logistic model with fixed effects only and logistic mixed model is the model-based standard error but not the point estimation, while IPCW point estimators and variance estimators only use the information from the point estimation of fixed and random effects of the censoring models. In terms of estimating the average outcomes and their corresponding closed-form variances, both censoring models provided similar results for the simulation scenario described in Section 3. When one anticipates a correlation of the censoring indicators within a network component either due to prior knowledge or estimated in the study data, the mixed effects censoring model can provide the information on the correlations between censoring indicators, a measure of the association between the missingness of participant outcomes. Therefore, the selection of the censoring model is based on what information is needed and if studying the association of the censoring mechanism between participants is of substantive interest.

We demonstrated both IPCW estimators to be consistent and asymptotically normal. We also derived a consistent estimator of the asymptotic variance. The simulation study showed that the IPCW estimator performed well in finite samples given a large number ($>100$) of components in the network. With the proposed method, we developed an approach to quantify a social and biological spillover effects on the risk of HIV transmission in HIV risk networks among PWID \cite{friedmanSocialNet2001, friedman2014socially, nikolopoulos2016network}. In additional simulation scenario 1, we increased the complexity of dependency for the censoring indicators in a component and showed that the performance of both IPCW estimators were comparable in terms of finite-sample performance. The additional simulation scenario 2 compared the performance of the IPCW estimators using the observed TRIP network (10 components) or the 24 components TRIP network. The variances were conservative when using the 10 component network, which may be due to the small number of components and the larger variability of the component sizes. After further dividing the network into 24 components using community detection, the estimated variances were anti-conservative but closer to empirical standard errors which agreed with the simulation results in the main scenario of small number of components in a network. This method leverages information on participants lost to follow-up in the network study to address the selection bias from differential loss to follow-up due to study drop out and possible distortion of the network structure resulting from removing individuals who had missing outcomes as when a complete case analysis been conducted.

Using two different assumptions about the correlation of censoring mechanisms, we applied the IPCW estimators to assess the causal effects of community alerts on HIV risk behavior at six-month follow-up in TRIP. There were 79 isolates in TRIP network which might be relevant for direct and overall effects. We focus on evaluating spillover effects which is not possible for isolates, we removed 79 isolates for our analysis. We estimated the variances of each causal effect using 10 and 24 components in the TRIP network under both censoring mechanisms. The mixed effects censoring model did not detect any random effect of each component which led to identical results using logistic censoring model and mixed effects model. The possible reason for this result is that the connected component in the network may have discordant random effects which eliminated with each other. The estimated CIs were rather wide while considering the original TRIP network (10 components) which may be caused by the variability of component sizes. After further dividing the network into 24 components using fast greedy community detection, the estimated CIs were narrowed (Table \ref{tab:trip}) and most of the estimated causal effects turned into significant. These results were aligned with the additional simulation scenario 2 that ECPs were over-estimated on 10 components TRIP network and slightly under-estimated on 24 components network. However, in practice, we are not able to measure the efficiency in estimating the CIs when further dividing the network into smaller components. In this study, community alerts were protective, and there were possible spillover effects from neighbors to an unexposed participant. The spillover effects were between $-0.06$ and $-0.21$ under allocation strategies 25\%, 50\%, and 75\%. 


There are some interesting future research directions for this work. One prominent direction is related to the choice and correct specification of the censoring model. Although the addition simulation scenario of varying variances of random effects in the censoring weight model showed comparable results in this study setting, the results might be different in other settings; for example, in the studies of participants grouped in families, when the dependency of censoring indicators between closely connected participants should be considered to improve the estimation of the variance \cite{wooldrige2007}. In addition, more censoring mechanisms should be considered such as missing not at random. For example,  a graphical models called "missingness graphs" are causal directed acyclic graphs that can be used to analyze missing not at random information and provides an effective way of representing the missingness mechanisms and potentially adjusting the estimators of causal parameters in interest from partially observed data \cite{mnar2011, mnar2013, mnar2018}. The consistency of the IPCW estimators requires correct specification of censoring and exposure weight models, so it is important to assess the model fit and conduct sensitivity analysis in the application of the proposed method \cite{misspecified}. It is worth noting that, due to the large sample properties, we require the network has fairly large number of components. The other direction is related to evaluating the accuracy of the variance estimator when the sample size or number of components is small \cite{minsamplesize}. Based on the simulation study, the confidence interval coverage levels can be below the nominal level in a network with a small number of network components or above the nominal level when the variability of the size of components is high. Future research could include developing a methodology that has reasonable finite-sample performance when the network has a small number of components or small number of participants. Additionally, the random effect in the same component was assumed homogeneous in this study. However, this may not hold when some of components in the network have large size. Future work could involve a revised M-estimation procedure for the variance to use individually-weighted estimators that are consistent when there are varying component sizes, extending results from a two-stage randomized trial to this network setting \cite{basse2018}.

\section*{Acknowledgements}
These findings are presented on behalf of the Transmission Reduction Intervention Project (TRIP). We would like to thank all of the TRIP investigators, data management teams, and participants who contributed to this project. The project described was supported grant DP2DA046856 by the Avenir Award Program for Research on Substance Abuse and HIV/AIDS (DP2) from National Institute on Drug Abuse of the National Institutes of Health, the National Institute on Drug Abuse of the National Institutes of Health award number DP1 DA034989, which funded Preventing HIV Transmission by Recently-Infected Drug Users, the National Institute on Drug Abuse of the National Institutes of Health award number P30DA011041 which supported the Center for Drug Use and HIV Research, and the National Institute of Allergy and Infectious Diseases of the National Institutes of Health award number R01 AI085073 Causal Inference in Infectious Disease Prevention Studies. Special thanks to Dr. Samuel R. Friedman (Department of Population Health, NYU Grossman School of Medicine) for his generous support. The content is solely the responsibility of the authors and does not necessarily represent the official views of the National Institutes of Health.

\bibliographystyle{plain} \bibliography{network}

\appendix
\section{Further details of Propositions}\label{appA}
\setcounter{theorem}{0}
\begin{theorem}
If the propensity scores and censoring weights are known, then the IPCW estimator is unbiased. $E[\widehat{Y}^{IPCW}(a, \alpha)]=\bar{y}(a, \alpha)$ and $E[\widehat{Y}^{IPCW}( \alpha)]=\bar{y}(\alpha)$.
\end{theorem}
\begin{proof}
The expected values of the estimator can be straightforwardly derived.
\begin{align}
    &E[\widehat{Y}^{ICPW}(a, \alpha)]=\frac{1}{n}\sum_{i=1}^n E\big[\frac{y_i(A_i, A_{\N_i})I(C_i=0) I(A_i=a)\pi(A_{\N_i};\alpha)}{f(A_i, A_{\N_i}|Z_i, Z_{\N_i})S_C(C_i|Z_i, Z_{\N_i}, A_i, A_{\N_i})}\big]\notag\\
    =&\frac{1}{n}\sum_{i=1}^n \sum_{a_i, a_{\N_i}} \sum_{c_i=1, 0}\frac{y_i(a_i, a_{\N_i})I(c_i=0)\cdot I(a_i=a)\pi(a_{\N_i};\alpha)}{f(a_i, a_{\N_i}|Z_i, Z_{\N_i})S_C(c_i|Z_i, Z_{\N_i}, A_i, A_{\N_i})}f(a_i, a_{\N_i}|Z_i, Z_{\N_i})S_C(c_i|Z_i, Z_{\N_i}, A_i, A_{\N_i})\notag\\
    =&\frac{1}{n}\sum_{i=1}^n \sum_{ a_{\N_i}}y_i(a_i=a, a_{\N_i})\pi(a_{\N_i};\alpha)\notag
\end{align}
The derivation of expected value of the IPCW marginal estimator is similar.
\end{proof}

\begin{theorem}
 Under suitable regularity conditions and due to the unbiased estimating equations, $\sqrt{m}(\hat{\theta}-\theta)$ converges in distribution to $N(0, \Sigma_m)$ as $m\rightarrow\infty$ where the variance matrix is given by $$\Sigma_m=\frac{1}{m}A^{-1}(\theta)B(\theta)(A^{-1}(\theta))^T.$$
\end{theorem}

Estimates $\hat{\gamma}$ and $\hat{\eta}$ that maximize the log likelihood are solutions to the score equations $$\sum_{\nu=1}^m\psi(Y_\nu, A_\nu, Z_\nu, C_\nu; \gamma)=0 \mbox{ and } \sum_{\nu=1}^m\psi(Y_\nu, A_\nu, Z_\nu, C_\nu; \eta)=0.$$ By M-estimation theory followed with Slutsky’s Theorem and Delta method as $m\rightarrow \infty$, $\hat{\theta}\xrightarrow{p}\theta$ and $\sqrt{m}(\hat{\theta}-\theta)$ converges in distribution to a multivariate normal distribution $N(0, \Sigma_m)$. The true parameter $\theta$ is defined as the solution to the equation $$\int \psi(Y_\nu, A_\nu, Z_\nu, C_\nu;\theta)dF_\nu(Y_\nu, A_\nu, Z_\nu, C_\nu)=0,$$ where $F$ is the cumulative distribution function of $(Y_\nu, A_\nu, Z_\nu, C_\nu)$. 

Replacing $A(\theta)$ and $B(\theta)$ with empirical estimators yields the empirical variance estimator $$\hat{\Sigma}_m=\frac{1}{m}A_m^{-1}(\hat{\theta})B_m(\hat{\theta})(A_m^{-1}(\hat{\theta}))^T$$ where $m$ is the number of components in the network. 
Let
$A_{11}(Y_\nu, A_\nu, Z_\nu, C_\nu)=\begin{pmatrix} 
\displaystyle\frac{\partial \psi(Y_\nu, A_\nu, Z_\nu, C_\nu; \gamma)}{\partial \gamma'}
\end{pmatrix}_{\gamma, \gamma'\in\Theta}$, $A_{22}(Y_\nu, A_\nu, Z_\nu, C_\nu)=\begin{pmatrix} 
\displaystyle\frac{\partial \psi(Y_\nu, A_\nu, Z_\nu, C_\nu; \eta)}{\partial \eta'}
\end{pmatrix}_{\eta, \eta'\in\Phi}$, $A_{31}(Y_\nu, A_\nu, Z_\nu, C_\nu; \alpha)=\begin{pmatrix} 
\displaystyle\frac{\partial \psi_0(Y_\nu, A_\nu, Z_\nu, C_\nu; \alpha, \theta_{0, \alpha})}{\partial \gamma}\\ \displaystyle\frac{\partial \psi_1(Y_\nu, A_\nu, Z_\nu, C_\nu; \alpha, \theta_{1, \alpha})}{\partial \gamma} \\
\displaystyle\frac{\partial \psi_2(Y_\nu, A_\nu, Z_\nu, C_\nu; \alpha, \theta_\alpha)}{\partial \gamma}\end{pmatrix}_{\gamma\in\Theta}$, and $A_{32}(Y_\nu, A_\nu, Z_\nu, C_\nu; \alpha)=\begin{pmatrix} 
\displaystyle\frac{\partial \psi_0(Y_\nu, A_\nu, Z_\nu, C_\nu; \alpha, \theta_{0, \alpha})}{\partial \eta}\\ \displaystyle\frac{\partial \psi_1(Y_\nu, A_\nu, Z_\nu, C_\nu; \alpha, \theta_{1, \alpha})}{\partial \eta} \\
\displaystyle\frac{\partial \psi_2(Y_\nu, A_\nu, Z_\nu, C_\nu; \alpha, \theta_\alpha)}{\partial \eta}\end{pmatrix}_{\eta\in\Psi}$, then $A_m(\hat{\theta})$ and $B_m(\hat{\theta})$ can be written as
$$A_m(\hat{\theta})=-\frac{1}{m}\sum_{\nu=1}^m\begin{pmatrix}A_{11}(Y_\nu, A_\nu, Z_\nu, C_\nu)&0&0\\0&A_{22}(Y_\nu, A_\nu, Z_\nu, C_\nu)& 0 \\A_{31}(Y_\nu, A_\nu, Z_\nu, C_\nu; \alpha) & A_{32}(Y_\nu, A_\nu, Z_\nu, C_\nu; \alpha)&-I_{3\times 3}\end{pmatrix},$$ and $$B_m(\hat{\theta})=\frac{1}{m}\sum_{\nu=1}^m\psi(Y_\nu, A_\nu, Z_\nu, C_\nu; \alpha, \hat{\theta})\psi(Y_\nu, A_\nu, Z_\nu, C_\nu; \alpha, \hat{\theta})^T.$$

\newpage
\section{Simulation results}\label{appB}
We included the simulation results for networks with 10, 50, and 100 components. We compared the true values and the estimates of the inverse probability censoring weighted estimator of the population average potential outcomes under allocation strategies 25\%, 50\%, and 75\%. Bias is the average of true values minus the estimated values. Empirical standard errors (ESE) is the standard deviation of estimated means. Empirical coverage probability (ECP) are the proportion of simulations where the true parameters fall into the 95\% confidence intervals among the 1000 simulations. Note that the true values are slightly different between the two censoring mechanisms due to the different random samples drawn for each simulation scenario.

\begin{table}[H]
  \centering
  \caption{Simulation results of IPCW estimators using logistic regression censoring model (left) and mixed effect censoring model (right) under allocation strategies 25\%, 50\%, and 75\% for network with 10 components}
    \begin{tabular}{l|rrrrr|rrrrr}
    \toprule
       & \multicolumn{5}{c|}{Logistic regression} & \multicolumn{5}{c}{Mixed effects } \\
        & \multicolumn{1}{c}{True} & \multicolumn{1}{c}{Bias} & \multicolumn{1}{c}{ESE} & \multicolumn{1}{c}{ASE} & \multicolumn{1}{c|}{ECP} & \multicolumn{1}{c}{True} & \multicolumn{1}{c}{Bias} & \multicolumn{1}{c}{ESE} & \multicolumn{1}{c}{ASE} & \multicolumn{1}{c}{ECP} \\
    \midrule
    \midrule
    Y(1, 0.25) & 0.2485 & -0.0132 & 0.206 & 0.143 & 0.79  & 0.2484 & 0.0061 & 0.197 & 0.119 & 0.69 \\
    Y(1, 0.5) & 0.2257 & -0.0097 & 0.131 & 0.102 & 0.87  & 0.2273 & 0.0075 & 0.130  & 0.082 & 0.79 \\
    Y(1, 0.75) & 0.2047 & -0.0067 & 0.148 & 0.104 & 0.80   & 0.2069 & 0.0066 & 0.132 & 0.083 & 0.72 \\
    Y(0, 0.25) & 0.2276 & -0.0127 & 0.162 & 0.114 & 0.81  & 0.2282 & 0.0045 & 0.147 & 0.121 & 0.73 \\
     Y(0, 0.5) & 0.2741 & -0.0080 & 0.140  & 0.110  & 0.89  & 0.2733 & 0.0151 & 0.112 & 0.105 & 0.80 \\
    Y(0, 0.75) & 0.3252 & -0.0099 & 0.219 & 0.159 & 0.80   & 0.3230 & 0.0228 & 0.180  & 0.128 & 0.74 \\
    Y(0.25) & 0.2328 & -0.0128 & 0.135 & 0.104 & 0.86  & 0.2332 & 0.0049 & 0.121 & 0.103 & 0.78 \\
    Y(0.5) & 0.2499 & -0.0089 & 0.101 & 0.087 & 0.95  & 0.2503 & 0.0113 & 0.085 & 0.074 & 0.84 \\
    Y(0.75) & 0.2349 & -0.0075 & 0.127 & 0.098 & 0.84  & 0.2359 & 0.0107 & 0.109 & 0.074 & 0.75 \\
    \hline
    \end{tabular}%
    
    \noindent{\footnotesize ESE = empirical standard error; ASE = average estimated standard error; ECP = empirical coverage probability.}
  \label{tab:addlabel}%
\end{table}%

\begin{table}[H]
  \centering
  \caption{Simulation results of IPCW estimators using logistic regression censoring model (left) and mixed effects censoring model (right) under allocation strategies 25\%, 50\%, and 75\% for network with 50 components}
    \begin{tabular}{l|rrrrr|rrrrr}
    \toprule
      & \multicolumn{5}{c|}{Logistic regression} & \multicolumn{5}{c}{Mixed effects } \\
        & \multicolumn{1}{c}{True} & \multicolumn{1}{c}{Bias} & \multicolumn{1}{c}{ESE} & \multicolumn{1}{c}{ASE} & \multicolumn{1}{c|}{ECP} & \multicolumn{1}{c}{True} & \multicolumn{1}{c}{Bias} & \multicolumn{1}{c}{ESE} & \multicolumn{1}{c}{ASE} & \multicolumn{1}{c}{ECP} \\
    \midrule
    \midrule
    Y(1, 0.25) & 0.2490 & 0.0045 & 0.073 & 0.068 & 0.87  & 0.2487 & 0.0057 & 0.077 & 0.068 & 0.87 \\
    Y(1, 0.5) & 0.2266 & 0.0030 & 0.044 & 0.043 & 0.92  & 0.2266 & 0.0045 & 0.046 & 0.044 & 0.91 \\
    Y(1, 0.75) & 0.2055 & 0.0010 & 0.055 & 0.058 & 0.89  & 0.2057 & 0.0008 & 0.060  & 0.053 & 0.87 \\
     Y(0, 0.25) & 0.2277 & -0.0006 & 0.053 & 0.050  & 0.92  & 0.2277 & 0.0015 & 0.052 & 0.050  & 0.91 \\
     Y(0, 0.5) & 0.2740 & 0.0018 & 0.044 & 0.045 & 0.93  & 0.2743 & 0.0046 & 0.043 & 0.044 & 0.93 \\
     Y(0, 0.75) & 0.3243 & -0.0015 & 0.083 & 0.074 & 0.89  & 0.3551 & 0.0346 & 0.080  & 0.074 & 0.89 \\
    Y(0.25) & 0.2330 & 0.0007 & 0.044 & 0.043 & 0.92  & 0.2330 & 0.0026 & 0.043 & 0.042 & 0.91 \\
    Y(0.5) & 0.2503 & 0.0024 & 0.031 & 0.032 & 0.93  & 0.2504 & 0.0046 & 0.031 & 0.032 & 0.94 \\
    Y(0.75) & 0.2352 & 0.0004 & 0.045 & 0.049 & 0.90   & 0.2355 & 0.0017 & 0.048 & 0.045 & 0.90 \\
    \hline
    \end{tabular}%
    
    \noindent{\footnotesize ESE = empirical standard error; ASE = average estimated standard error; ECP = empirical coverage probability.}
  \label{tab:addlabel}%
\end{table}%

\begin{table}[H]
  \centering
  \caption{Simulation results of IPCW estimators using logistic regression censoring model (left) and mixed effects censoring model (right) under allocation strategies 25\%, 50\%, and 75\% for network with 100 components}
    \begin{tabular}{l|rrrrr|rrrrr}
    \toprule
          & \multicolumn{5}{c|}{Logistic regression} & \multicolumn{5}{c}{Mixed effects} \\
          & True  & \multicolumn{1}{l}{Bias} & \multicolumn{1}{l}{ESE} & \multicolumn{1}{l}{ASE} & \multicolumn{1}{l|}{ECP} & True  & \multicolumn{1}{l}{Bias} & \multicolumn{1}{l}{ESE} & \multicolumn{1}{l}{ASE} & \multicolumn{1}{l}{ECP} \\
    \midrule
    \midrule
    Y(1, 0.25) & 0.2487 & 0.0022 & 0.053 & 0.050  & 0.90   & 0.2484 & 0.0010 & 0.052 & 0.051 & 0.91 \\
    Y(1, 0.5) & 0.2265 & 0.0002 & 0.031 & 0.031 & 0.94  & 0.2278 & -0.0014 & 0.037 & 0.039 & 0.94 \\
    Y(1, 0.75) & 0.2053 & -0.0049 & 0.040  & 0.039 & 0.92  & 0.2330 & -0.0008 & 0.030  & 0.033 & 0.94 \\
    Y(0, 0.25) & 0.2274 & -0.0039 & 0.037 & 0.038 & 0.94  & 0.2262 & 0.0006 & 0.032 & 0.032 & 0.94 \\
    Y(0, 0.5) & 0.2740 & 0.0010 & 0.030  & 0.031 & 0.96  & 0.2745 & 0.0028 & 0.031 & 0.031 & 0.94 \\
    Y(0, 0.75) & 0.3248 & 0.0033 & 0.054 & 0.053 & 0.91  & 0.2504 & 0.0017 & 0.022 & 0.023 & 0.96 \\
    Y(0.25) & 0.2327 & -0.0023 & 0.030  & 0.032 & 0.95  & 0.2054 & -0.0024 & 0.040  & 0.041 & 0.91 \\
    Y(0.5) & 0.2502 & 0.0006 & 0.021 & 0.023 & 0.95  & 0.3250 & 0.0036 & 0.055 & 0.053 & 0.91 \\
    Y(0.75) & 0.2352 & -0.0028 & 0.032 & 0.033 & 0.94  & 0.2353 & -0.0009 & 0.033 & 0.034 & 0.93 \\
    \bottomrule
    \end{tabular}%
    
    \noindent{\footnotesize ESE = empirical standard error; ASE = average estimated standard error; ECP = empirical coverage probability.}
  \label{tab:result_100}%
\end{table}%

\begin{table}[H]
  \centering
  \caption{Simulation results of IPCW estimators using logistic regression censoring model (left) and mixed effects censoring model (right) under allocation strategies 25\%, 50\%, and 75\% for a network with 200 components $^{a}$}
    \begin{tabular}{l|rrrrr|rrrrr}
    \toprule
          & \multicolumn{5}{c|}{Logistic regression} & \multicolumn{5}{c}{Mixed effects} \\
          & True  & \multicolumn{1}{l}{Bias} & \multicolumn{1}{l}{ESE} & \multicolumn{1}{l}{ASE} & \multicolumn{1}{l|}{ECP} & True  & \multicolumn{1}{l}{Bias} & \multicolumn{1}{l}{ESE} & \multicolumn{1}{l}{ASE} & \multicolumn{1}{l}{ECP} \\
    \midrule
    \midrule
    Y(1, 0.25) & 0.2486 & -0.0007 & 0.038 & 0.037 & 0.93  & 0.2485 & 0.0028 & 0.037 & 0.036 & 0.92 \\
    Y(1, 0.5) & 0.2264 & -0.0001 & 0.022 & 0.023 & 0.95  & 0.2263 & 0.0024 & 0.022 & 0.022 & 0.94 \\
    Y(1, 0.75) & 0.2053 & -0.0021 & 0.027 & 0.029 & 0.95  & 0.2052 & -0.0018 & 0.028 & 0.028 & 0.93 \\
    Y(0, 0.25) & 0.2275 & -0.0048 & 0.025 & 0.031 & 0.95  & 0.2281 & -0.0016 & 0.026 & 0.026 & 0.95 \\
    Y(0, 0.5) & 0.2745 & 0.0008 & 0.020  & 0.022 & 0.97  & 0.2747 & 0.0033 & 0.021 & 0.022 & 0.96 \\
    Y(0, 0.75) & 0.3250 & 0.0027 & 0.037 & 0.038 & 0.95  & 0.3252 & 0.0049 & 0.039 & 0.038 & 0.93 \\
    Y(0.25) & 0.2328 & -0.0037 & 0.020  & 0.025 & 0.97  & 0.2332 & -0.0005 & 0.021 & 0.022 & 0.95 \\
    Y(0.5) & 0.2505 & 0.0004 & 0.015 & 0.017 & 0.97  & 0.2505 & 0.0029 & 0.015 & 0.016 & 0.95 \\
    Y(0.75) & 0.2352 & -0.0009 & 0.022 & 0.024 & 0.95  & 0.2352 & -0.0001 & 0.023 & 0.023 & 0.93 \\
    \bottomrule
    \end{tabular}%
  \label{tab:result_200}%
  
  \noindent{\footnotesize$^{a}$ ESE = empirical standard error; ASE = average estimated standard error; ECP = empirical coverage probability.}
  
  \noindent{\footnotesize Note: The true values are slightly different between two censoring mechanisms due to the different random samples that were produced for each simulation. }
\end{table}%

\newpage
\section{Simulations: Scenario that varies variance of random effect in censoring weight model}
We conducted additional scenarios that varies the variance of the random effect in the censoring model. We used the simulated regular network with degree 4 and 100 components to compare the performance of the average outcome estimators using both censoring models $\widehat{Y}(1, \alpha)$, $\widehat{Y}(0, \alpha)$, and $\widehat{Y}(\alpha)$ where the censoring indicators are generated by 
$$C_i\sim\mbox{Bern}(\mbox{logit}^{-1}(-3+2\cdot Z_i+\rho_k))$$ where $\rho_k\sim N(0, r^2)$, $r=0.1, 0.2, 0.3, 0.4, 0.5$. The following figure compares the empirical coverage probability among 1000 simulations. 

\begin{figure}[H]
    \centering
    \includegraphics[scale=0.5]{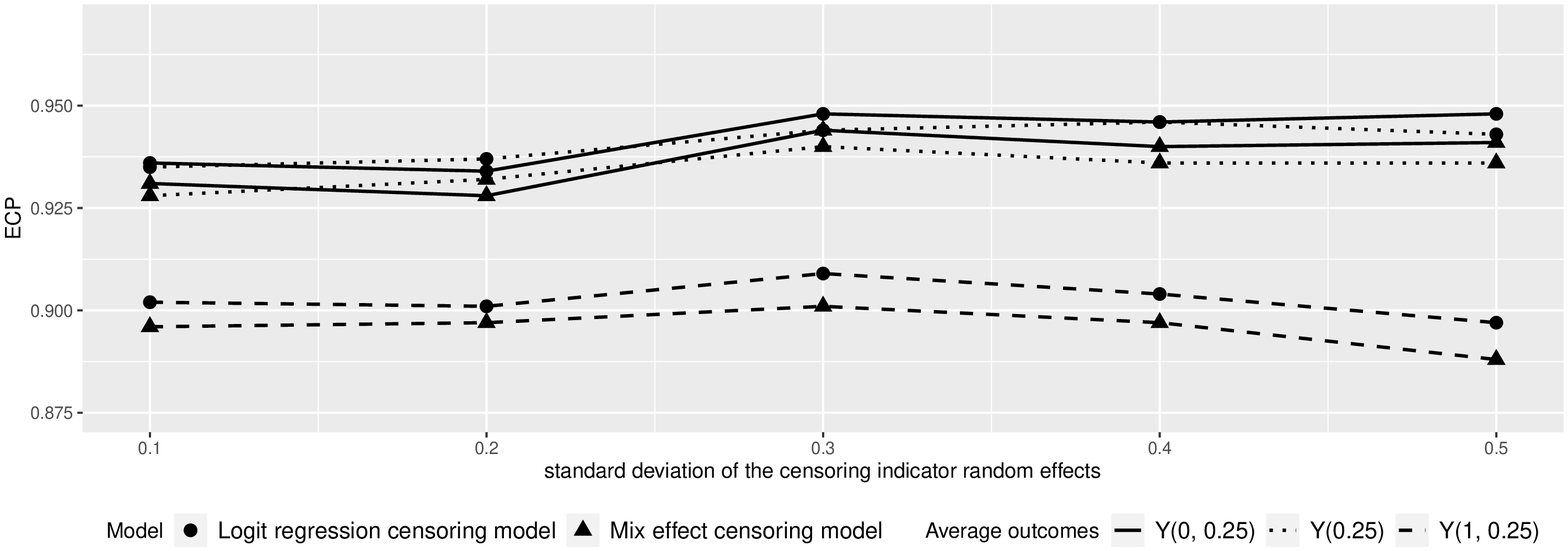}
    \includegraphics[scale=0.5]{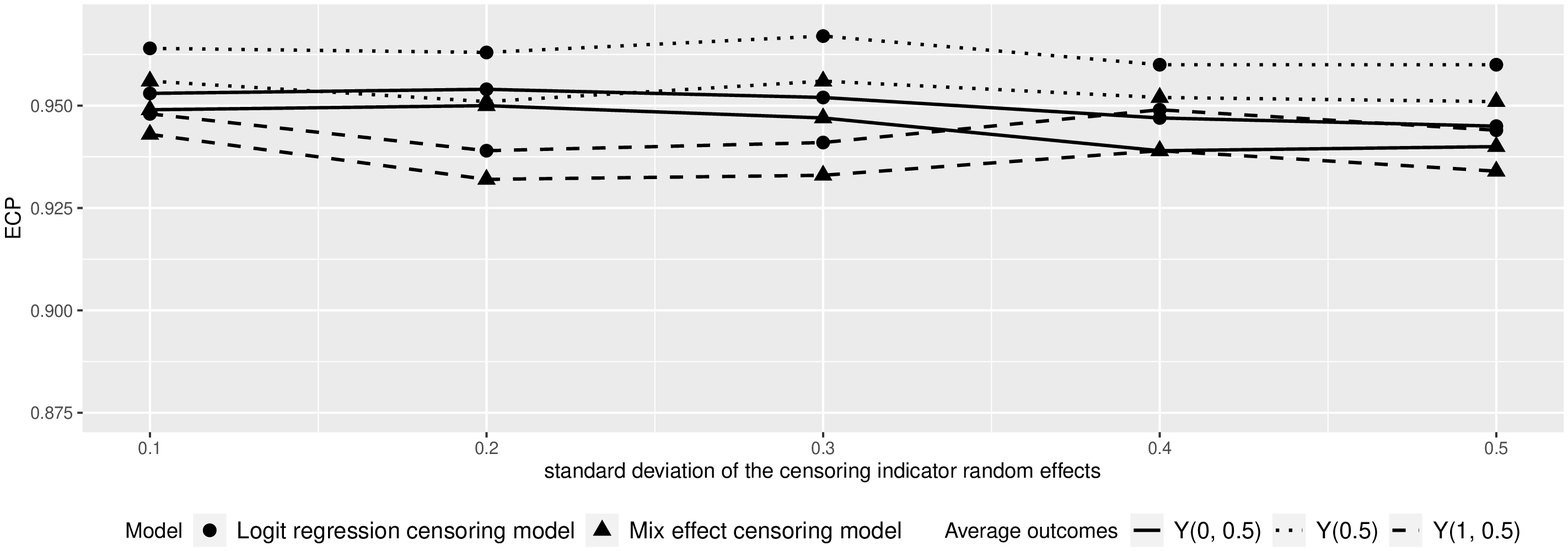}
    \includegraphics[scale=0.5]{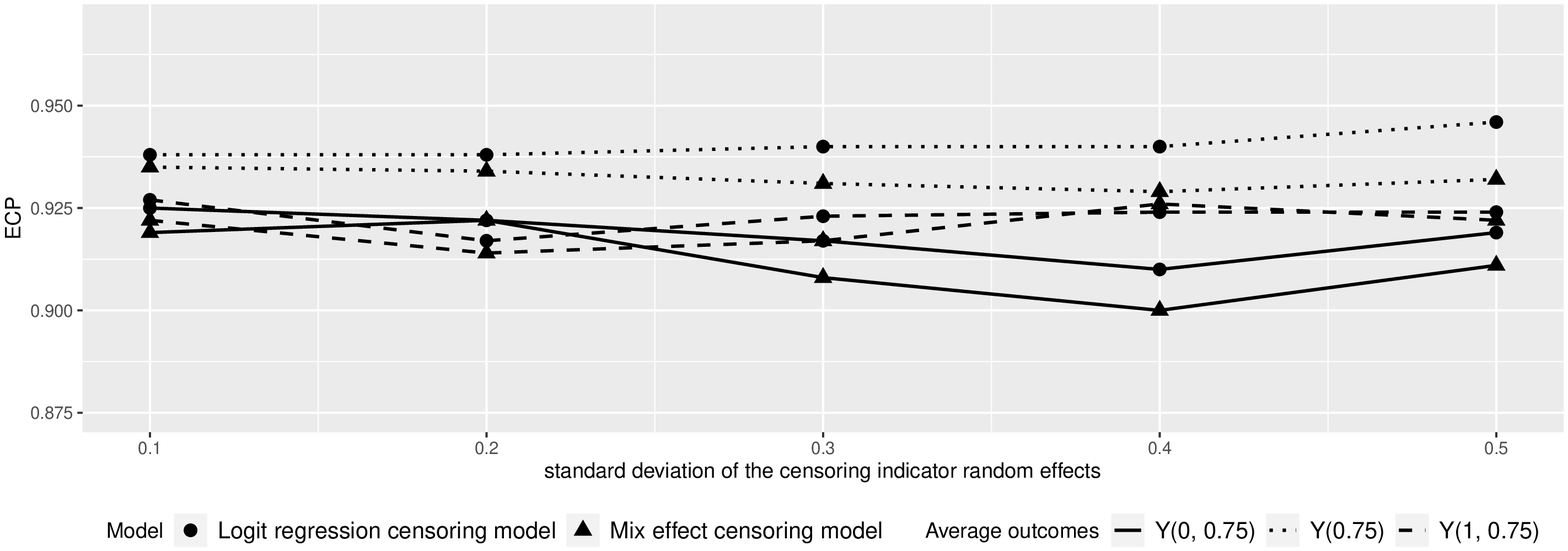}
    \caption{The empirical coverage probability (ECP) of the average outcomes at allocation strategies 25\% (top), 50\% (middle), and 75\% (bottom) using logistic regression and mixed effects censoring models when the standard deviations of censoring indicator random effects vary from 0.1 to 0.5.}
    \label{fig:my_label}
\end{figure}

\newpage
\section{Community Alerts and HIV Risk Behavior in TRIP at 6-month}\label{appC}

We further divided TRIP network into 24 components (the component sizes were 2, 2, 2, 2, 2, 3, 4, 5, 5, 5, 5, 7, 7, 8, 10, 10, 13, 16, 20, 21, 22, 30, 36, 38) to improved the validity of the estimation of the asymptotic variance of the causal effect estimates. The complete cases TRIP network ($n=216$) were divided into 20 components. We applied the IPCW estimators to quantify the causal effects of community alerts and HIV risk behavior on the report of risk behavior at the six-month visit. The mixed-effects censoring model did not detect random effects in each component on the original TRIP network. Therefore, the results using either censoring model are identical. The following table summarized the point estimations of the risk difference and 95\% Wald-type confidence intervals. 

\begin{table}[htbp]
  \centering
  \caption{The risk difference (RD) and 95\% Wald-type confidence intervals (CI) of the effects on the original TRIP network (24 components) (left) and estimated using completed cases (20 components) (right).}
    \begin{tabular}{lccccc}
    \toprule
    \multirow{2}[2]{*}{Effects} & Coverage & \multicolumn{2}{c}{Censoring Model} & \multicolumn{2}{c}{Complete cases}\\
          & ($\alpha$, $\alpha'$) & RD  & 95\% CI & RD  & 95\% CI \\
         
    \midrule
    \midrule
     Direct & (25\%, 25\%) & -0.1228 & (-0.341, 0.095) & -0.0772 & (-0.444, 0.290)\\
     Direct & (50\%, 50\%) & -0.2299 & (-0.401,-0.059) & -0.1761 & (-0.518, 0.165)\\
     Direct & (75\%, 75\%) & -0.1741 & (-0.357, 0.009) &-0.2393	& (-0.495, 0.016) \\
    Indirect & (50\%, 25\%) & -0.0565 & (-0.129, 0.016) &-0.0404	 & (-0.097, 0.016) \\
    Indirect & (75\%, 50\%) & -0.1502 & (-0.272,-0.028) & -0.0818	 & (-0.150,-0.013) \\
    Indirect & (75\%, 25\%) & -0.2066 & (-0.381,-0.032)& -0.1222	& (-0.238,-0.006)\\
     Total & (50\%, 25\%) & -0.2864 & (-0.494,-0.079) & -0.2165 &	(-0.531, 0.098) \\
     Total & (75\%, 50\%) & -0.3243 & (-0.503,-0.146) &-0.3211 & (-0.545,-0.097) \\
     Total & (75\%, 25\%) & -0.3808 & (-0.583,-0.179) & -0.3615 & (-0.557,-0.166)\\
    Overall & (50\%, 25\%) & -0.1407 & (-0.224,-0.057) & -0.1091 & (-0.174,-0.044) \\
    Overall & (75\%, 50\%) & -0.1658 & (-0.243,-0.089) & -0.1732 & (-0.300,-0.046) \\
    Overall & (75\%, 25\%) & -0.3065 & (-0.452,-0.161) & -0.2824 & (-0.414,-0.151)\\
    \bottomrule
    \end{tabular}%
  \label{tab:trip}%
\end{table}%

\end{document}